\pdfoutput=1
\documentclass[prl,twocolumn,preprintnumbers,nofootinbib,superscriptaddress,10pt]{revtex4-1}
\usepackage[dvips]{graphicx}

\usepackage{amsmath,amsthm,amssymb,multirow,psfrag}
\usepackage[normalem]{ulem}
\usepackage{epsfig}
\usepackage{color}
\usepackage{hyperref}
\graphicspath{{./Figures/}}

\begin{document}

\def\lsim{\mathrel{\rlap{\lower4pt\hbox{\hskip1pt$\sim$}}
  \raise1pt\hbox{$<$}}}
\def\gsim{\mathrel{\rlap{\lower4pt\hbox{\hskip1pt$\sim$}}
  \raise1pt\hbox{$>$}}}
\newcommand{\vev}[1]{ \left\langle {#1} \right\rangle }
\newcommand{\bra}[1]{ \langle {#1} | }
\newcommand{\ket}[1]{ | {#1} \rangle }
\newcommand{\ev}{ {\rm eV} }
\newcommand{\kev}{{\rm keV}}
\newcommand{\Mev}{{\rm MeV}}
\newcommand{\Gev}{{\rm GeV}}
\newcommand{\Tev}{{\rm TeV}}
\newcommand{\mw}{$M_{W}$}
\newcommand{\Ft}{F_{T}}
\newcommand{\Zparity}{\mathbb{Z}_2}
\newcommand{\BLambda}{\boldsymbol{\lambda}}
\newcommand{\met}{\;\not\!\!\!{E}_T}
\newcommand{\beq}{\begin{equation}}
\newcommand{\eeq}{\end{equation}}
\newcommand{\bea}{\begin{eqnarray}}
\newcommand{\eea}{\end{eqnarray}}
\newcommand{\nn}{\nonumber\\}
\newcommand{\gev}{{\mathrm GeV}}
\newcommand{\hc}{\mathrm{h.c.}}
\newcommand{\eps}{\epsilon}
\newcommand{\bwt}{\begin{widetext}}
\newcommand{\ewt}{\end{widetext}}
\newcommand{\draftnote}[1]{{\bf\color{blue} #1}}
\newcommand{\vecsigma}{ \mbox{\boldmath $\sigma$}}
\newcommand{\vecp}{\mathbf p}
\newcommand{\vecq}{\mathbf q}
\newcommand{\veck}{\mathbf k}
\newcommand{\vecr}{\mathbf r}

\newcommand{\Mpl}{M_{\rm Pl}}
\newcommand{\TRH}{T_{\rm RH}}
\newcommand{\TCMB}{T_{\rm CMB}}
\newcommand{\xiend}{\xi_{\rm end}}
\newcommand{\Hend}{H_{\rm end}}
\newcommand{\HRH}{H_{\rm RH}}
\newcommand{\aRH}{a_{\rm RH}}
\newcommand{\aend}{a_{\rm end}}
\newcommand{\amax}{a_{\rm max}}
\newcommand{\abar}{a_{\rm bar}}
\newcommand{\kend}{k_{\rm end}}
\newcommand{\kmax}{k_{\rm max}}
\newcommand{\xend}{x_{\rm end}}
\newcommand{\kast}{k_\ast}
\newcommand{\rhoAend}{\rho^{\rm{end}}_E}
\newcommand{\rhoIend}{\rho^{\rm{end}}_I}
\newcommand{\sigend}{\overline{\sigma}_{\rm{end}}}
\newcommand{\sigRH}{\overline{\sigma}_{\rm{RH}}}
\newcommand{\rhochiend}{\rho^{\rm{end}}_\chi}
\newcommand{\mbchi}{\overline{m}_\chi}
\newcommand{\sigb}{\overline{\sigma}}
\newcommand{\TbA}{\overline{T}_A}
\newcommand{\Tbchi}{\overline{T}_\chi}

\newcommand{\cO}{{\cal O}}
\newcommand{\cL}{{\cal L}}
\newcommand{\cM}{{\cal M}}

\newcommand{\fref}[1]{Fig.~\ref{fig:#1}} 
\newcommand{\eref}[1]{Eq.~\eqref{eq:#1}} 
\newcommand{\aref}[1]{Appendix~\ref{app:#1}}
\newcommand{\sref}[1]{Section~\ref{sec:#1}}
\newcommand{\tref}[1]{Table~\ref{tab:#1}}

\newcommand{\LU}[1]{\textcolor{red}{[LU: #1]}}

\title{\LARGE{{\bf Schwinger dark matter production}}}

\author{\bf {Mar Bastero-Gil}}
\email{mbg@ugr.es}
\affiliation{\normalsize\it
Departamento de F\'{i}sica Te\'{o}rica y del Cosmos and CAFPE,~Universidad de Granada, Campus de Fuentenueva, E-18071 Granada, Spain}

\author{\bf{Paulo B. Ferraz}}
\email{paulo.ferraz@student.uc.pt}
\affiliation{\normalsize\it
Departamento de F\'{i}sica Te\'{o}rica y del Cosmos and CAFPE,~Universidad de Granada, Campus de Fuentenueva, E-18071 Granada, Spain}
\affiliation{\normalsize\it
Universidad de Coimbra, Faculdade de Ciências e Tecnologia da Universidade
de Coimbra and CFisUC, Rua Larga, 3004-516 Coimbra, Portugal}

\author{\bf{Lorenzo Ubaldi}}
\email{lorenzo.ubaldi@ijs.si}
\affiliation{\normalsize\it
Jo\v{z}ef Stefan Institute, Jamova 39, 1000 Ljubljana, Slovenia}
\affiliation{\normalsize\it
Institute for Fundamental Physics of the Universe (IFPU), \\ Via Beirut 2, 34014 Trieste, Italy}

\author{\bf{Roberto Vega-Morales}}
\email{rvegamorales@ugr.es\\}
\affiliation{\normalsize\it
Departamento de F\'{i}sica Te\'{o}rica y del Cosmos and CAFPE,~Universidad de Granada, Campus de Fuentenueva, E-18071 Granada, Spain}

\preprint{UG-FT 329-23,~CAFPE 199-23,~SISSA 44/2020/FISI,~CA21106}
\begin{abstract}
Building on recently constructed inflationary vector dark matter production mechanisms as well as studies of magnetogenesis, we show that an inflationary \emph{dark} Schwinger mechanism can generate the observed dark matter relic abundance for `dark electron' masses as light as $\sim 0.1$\,eV and as heavy as $10^{12}$\,GeV.~The dark matter can interact very weakly via the exchange of light dark photons with a power spectrum which is peaked at very small scales, thus evading isocurvature constraints.~This mechanism is viable even when (purely) gravitational particle production is negligible.~Thus dark matter can be produced solely via the Schwinger effect during inflation including for light masses. 

\end{abstract}
\maketitle

\section{Introduction} \label{sec:intro} 

The nature and production mechanism of dark matter is still unknown.~Significant research effort has been put into candidates like weakly interacting massive particles, axions, sterile neutrinos, but we have seen no experimental evidence for them yet~\cite{Bertone:2018xtm}.~This motivates studying alternative candidates as well.~One class which has received considerable attention in the recent past is that of dark photon dark matter, produced via non-thermal mechanisms~\cite{Graham:2015rva, Agrawal:2018vin, Dror:2018pdh, Co:2018lka,  Bastero-Gil:2018uel, Long:2019lwl, McDermott:2019lch, Nakai:2020cfw,Ahmed:2020fhc,Kolb:2020fwh, Salehian:2020asa, Bastero-Gil:2021wsf, Gorghetto:2022sue, Sato:2022jya, Redi:2022zkt, Barrie:2022mma} where the dark photon is associated with a dark $U(1)_D$ abelian gauge theory.~At some point before matter radiation equality the dark photon must become massive, even if very light, in order to provide a cold (non-relativistic) dark matter candidate.~However, it has been argued~\cite{East:2022rsi} that the formation of vortices in these scenarios spoils the possibility of the dark photon being dark matter\footnote{However, these constraints can be softened in some dark photon dark matter scenarios~\cite{Cyncynates:2023zwj}.}.~One can consider adding to the dark sector scalar and/or fermion fields charged under the $U(1)_D$, of which the dark photon is the gauge boson.~Utilizing one of the dark photon production mechanisms above, all that is needed is a way to transfer the energy density from the dark photon to the `dark electrons'\footnote{We will generically refer to both scalars and fermions charged under the dark  $U(1)_D$ as dark electrons.} which can then constitute the dark matter. 

Such a `dark QED' setup was studied in~\cite{Arvanitaki:2021qlj} which utilized the mechanism of~\cite{Graham:2015rva} to generate a dark electric field composed of the longitudinal mode of the (massive) dark photon which is produced via inflationary fluctuations.~While the modes are produced during inflation, the cold vector condensate, which can be treated as a classical dark electric field, only forms in the radiation dominated epoch when the vector momentum modes re-enter the horizon~\cite{Graham:2015rva}.~Since the dark electric field which is produced is not very large compared to the Hubble scale, this mechanism requires efficient transfer of the dark sector energy density from the dark photon to the dark electrons which requires the dark sector to thermalize in order to produce enough dark matter.~It was shown~\cite{Arvanitaki:2021qlj} how via various dark QED processes such as electromagnetic cascades, and plasma dynamics, as well as a subdominant contribution from Schwinger pair production, the dark sector can thermalize with itself while remaining decoupled from the visible thermal bath.~In this case the amount of dark electrons today is determined by the interplay between the various thermalization processes in the dark sector and one obtains a relic abundance which matches the one observed for dark matter in the mass range from $50$\,MeV to $30$\,TeV.~In order to achieve thermalization the dark $U(1)_D$ gauge coupling cannot be too small, roughly $\gtrsim 10^{-3}$.~Furthermore, as in~\cite{Graham:2015rva}, the dark photon must be massive for the whole of its cosmological evolution including during inflation.

We consider a similar dark QED scenario with the same field content as in~\cite{Arvanitaki:2021qlj}, but instead where the dark electric field is generated by the mechanism proposed in~\cite{Bastero-Gil:2018uel,Bastero-Gil:2021wsf} which is based on an axion like coupling between the gauge field and the inflaton.~Compared to the longitudinal mode production mechanism~\cite{Graham:2015rva} discussed above, the mechanism utilized here produces much larger dark electric fields already during inflation via one of the \emph{transverse} vector modes, thus it also works in the massless dark photon case.~Once formed these background dark electric fields can then source Schwinger pair production of dark electrons during inflation.

At the end of inflation and the onset of reheating, which we assume instantaneous, we have an energy density in the dark sector which is dominated by the dark electric field along with a tiny seed of Schwinger produced dark electrons.~In the thermalization case discussed above, there is also Schwinger pair production, but any initial seed of dark electrons coming from Scwhinger production is washed out by the more dominant thermalization effects~\cite{Arvanitaki:2021qlj}.~Here we consider the regime where the dark sector does \emph{not} thermalize with itself which implies an upper bound on the dark gauge coupling.

In this case the only source of dark electrons is from inflationary dark Schwinger production.~After reheating Schwinger production ceases and at some point the dark electrons begin redshifting like non-relativistic matter, while the electric field continues redshifting like radiation, eventually coming to dominate the dark sector energy density after matter-radiation equality.~Thus, the relic abundance is set by the balance between inflationary Schwinger production and redshifting effects.

Following this evolution we show that Schwinger produced dark electrons can reproduce the observed dark matter relic abundance for masses potentially as light as $\sim 0.1$\,eV and as heavy as $\sim 10^{12}$\,GeV.~The dark gauge coupling can range from $\sim 10^{-9} - 1$ while the Hubble scale at the end of inflation can be in the range $\sim (100 - 10^{13})$\,GeV.~We expect the dark electrons to have a power spectrum peaked around the same scales as the dark photon which is at very small scales~\cite{Bastero-Gil:2021wsf}, thus evading constraints on isocurvature coming from measurements of the CMB.

The Schwinger effect in de-Sitter space has been explored in various studies~\cite{Frob:2014zka,Kobayashi:2014zza,Hayashinaka:2016qqn,Banyeres:2018aax,Domcke:2018eki,Stahl:2018idd,Sobol:2019xls,Domcke:2019qmm,Sobol:2020frh} in the context of magnetogenesis to explain the potential existence of primordial magnetic fields.~In these studies the conserved Schwinger current during inflation was computed for both scalars~\cite{Banyeres:2018aax} and fermions~\cite{Hayashinaka:2016qqn}.~We adapt these results to study the evolution of the energy density of the produced dark sector particles averaged over space in order to estimate the parameter space where we can obtain the correct dark matter abundance.

We also emphasize that conformal invariance must be broken during inflation to generate particles.~In inflationary Schwinger production the conformal symmetry breaking comes from the background electric field (which can be generated in various ways).~This breaking of conformal invariance ‘leaks’ into the matter sector via the Schwinger mechanism allowing for production of charged particles even in the absence of conformal breaking in the fermion and scalar sectors.~In this way even massless fermions or conformally coupled massless scalars, which would not be produced by purely gravitational particle production due to a changing metric~\cite{Ford:1986sy,Kuzmin:1998kk}, can still be produced during inflation via the Schwinger effect.

This paper is organized as follows.~We first give a brief review of Schwinger production during inflation and compute the dark sector energy density at the end of inflation.~We also review the electric field production mechanism based on an axion like coupling between the gauge field and the inflaton.~We then track the cosmic evolution of the dark sector and compute the relic abundance today.~Finally, we examine the viable dark matter parameter space of this mechanism before discussing potential future directions and concluding.

\section{Dark Schwinger production}\label{sec:schwinger}

In this section we first review the relevant aspects of the Schwinger mechanism needed for production of dark charged particles in the expanding universe.~Details of the results summarized and applied to dark  QED here can be found in~\cite{Kobayashi:2014zza,Hayashinaka:2016qqn,Banyeres:2018aax},
which studied the Schwinger effect in deSitter space, relevant for the period of inflation.

We consider a dark $U(1)_D$ gauge field during inflation with gauge coupling $g_D$ which couples to dark charged fermions and/or scalars which we label $\chi$.~We assume $\chi$ to have dark charge $Q_\chi = 1$ and to be a singlet under the Standard Model.~In analogy with standard QED,
\bea\label{eq:dQEDaction}
S&=&
\int d^4x \sqrt{-g}
\left(
-\frac{1}{4} F^{\mu\nu}F_{\mu\nu} +
\, \mathcal{L}_{\rm{charged}}(A_\mu,\chi) 
\right) ,~~
\eea 
where for charged scalars we have,
\bea \label{eq:lagscal}
\mathcal{L}_{\rm charged}^{\rm scalar} = -|D_\mu \chi|^2 - (m_\chi^2 + \xi_R R)|\chi |^2 \, ,
\eea
and for charged fermions,
\bea \label{eq:lagferm}
\mathcal{L}_{\rm charged}^{\rm fermion} = \bar\chi ( i \gamma^\mu D_\mu - m_\chi) \chi \, .
\eea
Here $F_{\mu\nu}= \partial_\mu A_\nu - \partial_\nu A_\mu$, while $D_\mu = \partial_\mu + i g_D A_\mu + \Gamma_\mu$, where $\Gamma_\mu$ is the spinor connection (which is zero in the scalar case) and $R$ is the Ricci
scalar.~We parametrize the background de-Sitter spacetime by the Friedmann-Robertson-Walker metric with $ds^2 = - dt^2 + a^2(t) d\vec{x}^2$ and the convention $\epsilon^{0123} = 1/\sqrt{-g}$.

In order to isolate the Schwinger production from the possible gravitational production~\cite{Ford:1986sy,Kuzmin:1998kk}, we consider the conformal
limit where $m_\chi / H \ll 1$ with $H$ the Hubble parameter during inflation and $\xi_R = 1/6$ for the scalar case in~\eref{lagscal}.~While we can neglect $m_\chi$ during inflation in this limit, at some point in its cosmic evolution $\chi$ must obtain a mass to eventually become non-relativistic, as required for cold dark matter.~We also assume for now a constant classical dark electric field during inflation and discuss in the next section how it can be generated.

The electric field has a direction, which we take to be along the $z$ axis, and will generate a current of charged particles via the Schwinger effect.~Calculation of the current is challenging and requires renormalization, but has been performed both for scalars~\cite{Kobayashi:2014zza} and fermions~\cite{Hayashinaka:2016qqn}
in $3+1$ de-Sitter space.~Following the notation of~\cite{Kobayashi:2014zza}, we can write the electric field in the $z$ direction as $E_i = a E \delta_{iz}$ with $E$ constant and the current as $\langle J_i \rangle = a J \delta_{iz}$.~In both
cases $a$ is the comoving scale factor which appears in the metric, and we refer to the quantities $E$ and $J$ as the physical electric field and physical current respectively.~We can then define the electrical conductivity, and its dimensionless counterpart, as
\bea
\sigma = \frac{J}{E} \, , \qquad  \bar \sigma \equiv \frac{\sigma}{H} \, .
\eea
As we'll see below, with the conductivity as input we can compute the energy density of the Schwinger produced charged particles during inflation in terms of the energy density of the dark electric field. 

To estimate the energy density $\rho_\chi$ of dark charged particles we assume they form a perfect fluid whose evolution during inflation we can describe as~\cite{Sobol:2019xls},
\bea\label{eq:rhochievo}
\dot{\rho}_{\chi} + nH\rho_{\chi} = \langle E \cdot J \rangle
= \sigma \langle E^2 \rangle .
\eea
Here $\langle E^2 \rangle$ represents the spatial average of the magnitude squared of the dark electric field,~We also have $n = 3$ for the non-relativistic case and $n = 4$ for the relativistic case while the overdot denotes a derivative with respect to cosmic time $t$.~This evolution equation can be derived from the energy momentum tensor and expresses energy conservation where $\langle E\cdot J \rangle$ acts as a source term for the $\chi$ energy density.~A corresponding term, with the opposite sign, appears in the evolution equation for the energy density of the dark electric field. This must be sourced during inflation, as we discuss in the next section, but is depleted by the $- \sigma \langle E^2 \rangle$ term, which is due to the Schwinger production\,\footnote{For these considerations to be sensible, $\langle E\cdot J \rangle$ must be positive, that is $\sigma > 0$. It was pointed out in~\cite{Hayashinaka:2016qqn} that the conductivity could turn negative in the limit of weak electric force and small mass of the charged particle.~However this effect is unlikely to be physical~\cite{Banyeres:2018aax}.~We believe this point deserves further investigation which we leave for ongoing work.~In this study we assume that the conductivity is always positive.}.\,We work in the small conductivity regime where,
\bea \label{eq:sigmassump}
0 < \bar \sigma \ll 1 \, ,
\eea
so we can neglect the depleting term in the evolution equation for the energy density of the dark electric field and treat~\eref{rhochievo} for $\rho_\chi$ as decoupled.~If $m_\chi /H$ is not small during inflation or the dark scalar has a non-conformal coupling to gravity there should be an additional term present in~\eref{rhochievo} due to gravitational particle production~\cite{Ford:1986sy,Kuzmin:1998kk}.
As mentioned above, we assume we are near the conformal limit where this production is negligible and leave a study of the conformal breaking case where $m_\chi /H > 1$ to a companion paper~\cite{Bastero-Gil:2023htv}.

During inflation it is convenient to rewrite~\eref{rhochievo} switching to conformal time where $d\tau = dt/a$,
\bea
\partial_{\tau}\rho_{\chi}-\frac{n}{\tau}\rho_{\chi} = 
- \overline{\sigma}E^2\frac{1}{\tau} ,
\eea
and we have used $\tau = -1/(aH)$.~Taking $\rho_\chi(\tau = \tau_0) = 0$ as the initial condition the solution is,
\bea\label{eq:rhochi}
\rho_{\chi}(\tau) =  \frac{\overline{\sigma}}{n}E^2
 \Big[1-\Big(\frac{\tau_0}{\tau}\Big)^{-n}\Big]  .
\eea
Note that $\tau_0/\tau_f = e^{N_e}$ with $N_e$ the number of e-folds.~So if Schwinger production starts just a few e-folds before the end of inflation, during which the electric field stays constant, then at the end of inflation we will have,
\bea\label{eq:rhochiapprox}
\rho_\chi^{\rm end} \equiv \rho_\chi(\tau_{\rm end}) \simeq \frac{\overline{\sigma}}{n} E^2 =   \frac{\overline{\sigma}}{n} 2 \rho_E^{\rm end}\, ,
\eea
where the energy density in the dark electric field (derived from the energy momentum tensor) at the end of inflation is given by~\cite{Bastero-Gil:2021wsf},
\bea\label{eq:rhoE}
\rhoAend = \frac{1}{2} E^2 .
\eea
We see that at the end of inflation the energy density of the dark charged particles is proportional to the dark electric field energy density, but suppressed by the (dimensionless) conductivity.~This result holds for both scalars and fermions and is independent of the source for the dark electric field.~The only requirement is that it be (approximately) constant within a Hubble patch and contribute \emph{some} amount to the energy density of the dark vector\footnote{We will use `dark vector' and `dark photon' interchangeably.}.~Thus any mechanism which generates a constant background dark electric field during inflation will also Schwinger produce some amount of dark particles.

In visible QED, after inflation and reheating the Standard Model (SM) plasma thermalizes and erases any initial seed of Schwinger produced electrons as well as any classical electric field generated during inflation.~Similar conclusions follow for a dark QED sector which thermalizes with itself~\cite{Arvanitaki:2021qlj}.~In this work we take the dark gauge coupling to be small enough that the dark QED sector never thermalizes within itself.~We also assume the dark sector is secluded from the visible one.~Thus the dark matter relic abundance is determined only by the inflationary Schwinger effect and redshifting. 

Whether the electric field can source Schwinger production after inflation is unclear since a calculation of the conductivity during radiation domination in an FRW background does not exist.~However, as we show in the Appendix, after reheating the electric field begins oscillating~\cite{Bastero-Gil:2021wsf} and redshifting like radiation and no longer accelerates the dark electrons efficiently.~We therefore expect Schwinger production to be suppressed after inflation though quantifying this requires further investigation.~In this study we assume any Schwinger production after reheating is negligible.

%

\subsection{Generating dark electric fields during inflation}\label{sec:Efield}

In order to form a classical dark electric field which can source Schwinger production during and after inflation, the vector bosons must be produced in a configuration with a very large occupation number and a coherence wavelength at least of order the horizon, $H^{-1}$.~Here we describe how to generate such a dark electric field during inflation.~Mechanisms for producing a vector field during inflation can be constructed in a number of ways.~One possibility is to consider a minimal coupling to the metric via a vector mass term~\cite{Graham:2015rva,Ahmed:2020fhc}.~Alternatively, one can consider non-minimal couplings between the inflaton and the gauge field.~The two most common such couplings are ones between (a function of) the inflaton and the kinetic term of the gauge field $I(\phi) FF$~\cite{Shakeri:2019mnt,Nakayama:2020rka,Nakai:2020cfw} or the dual field strength $I(\phi) F\widetilde{F}$~\cite{Bastero-Gil:2018uel,Bastero-Gil:2021wsf,Bastero-Gil:2022fme}.

Motived by, but not limited to models of axion inflation~\cite{Meer}, here we focus on the mechanism presented in~\cite{Bastero-Gil:2018uel,Bastero-Gil:2021wsf} which produces only \emph{one transverse} mode of the dark vector field and works even if the dark vector is massless.~This mechanism utilizes the $\phi F\widetilde{F}$ operator which introduces a source of conformal symmetry violation and a time dependence into the dark vector dispersion relation.~This induces a tachyonic instability and leads to exponential production (very high occupation number) of one of the transverse modes with a coherence length $l \gtrsim \Hend^{-1}$~\cite{Bastero-Gil:2018uel,Bastero-Gil:2021wsf}.~Thus we have a (approximately) constant classical electric field within each Hubble patch which can serve as the source for Schwinger production.

We add to the action in~\eref{dQEDaction} the Lagrangian,
\bea\label{eq:Lsource}
 \mathcal{L}_{\rm{source}}(A_\mu,\phi)&=&
- \frac{1}{2}\partial_{\mu}\phi\partial^{\mu}\phi - V(\phi)
- \frac{\alpha}{4f}\phi F_{\mu\nu}\widetilde{F}^{\mu\nu},~~~~~
\eea 
where $\phi$ is the inflaton field that drives inflation and $\widetilde F^{\mu\nu} = \epsilon^{\mu\nu\alpha\beta} F_{\alpha\beta}/2$ with $\epsilon^{\mu\nu\alpha\beta}$ the completely antisymmetric tensor.~We do not specify the inflaton potential $V(\phi)$ as its precise form is not crucial for the production mechanism considered here, but can affect the shape of the energy density spectrum as explored in detail in~\cite{Bastero-Gil:2021wsf}.~The dark vector can in principle have a mass, as long as it is small compared to the Hubble scale at the end of inflation.~We neglect the vector mass for the rest of this section, but we will consider it when examing the evolution of the energy densities after reheating.

The coupling $\alpha / (4f)$ introduces the scale $f$ and a source of conformal symmetry breaking into the theory which leads to the production of a large number of dark photons at the expense of the kinetic energy of the inflaton~\cite{Anber:2009ua, Barnaby:2011qe}.~We consider a regime where $\alpha / (4f)$ is small enough to ignore the back reaction effects on the inflaton due to dark photon production.~In doing so, we only need to write the equation of motion for $A_\mu$.~In the presence of dark charged particles, or equivalently of non-zero conductivity, the equation of motion is,
\bea\label{eq:EOMfin}
\ddot{A}_\pm + H(1 + \overline{\sigma} )\dot{A}_\pm+\Big(\frac{k^2}{a^2}
\mp 2\xi H \frac{k}{a}\Big)A_\pm=0,
\eea
where $\pm$ refers to the helicities of the transverse modes and we have defined the `instability parameter',
\bea \label{eq:xidef}
\xi \equiv \frac{\alpha \dot \phi}{2H f} = \sqrt{\frac{\epsilon}{2}} \frac{\alpha}{f} \Mpl  \, .
\eea
The `slow roll' parameter is defined as $\epsilon \equiv - \dot H/H^2$ where in single field inflation we have the relations,
\bea\label{eq:slowroll}
|\dot\phi| \approx  V'/3H,~~ 
\epsilon =  \frac{\dot \phi^2}{2H^2 M_{Pl}^2} .
\eea
Note the conductivity enters as a correction to the Hubble damping term in~\eref{EOMfin}.~As stated in~\eref{sigmassump}, we work in the regime $\bar\sigma \ll 1$, so we can neglect the effect of the conductivity on the transverse vector mode equation of motion.~Taking $\xi$ to be time-independent, the solution can be found  in terms of Whittaker functions using the WKB approximation as shown in~\cite{Bastero-Gil:2021wsf} for instance, to which we refer the reader for details.~From~\eref{EOMfin} we see that when $k/a \lesssim 2\xi H$, one of the transverse modes becomes tachyonic (which we take to be $A_+$) and exponentially enhanced which we can associate with particle production of the dark vector~\cite{Bastero-Gil:2018uel,Bastero-Gil:2021wsf}.

In our regime of interest, where we can neglect backreaction effects while still generating enough particle production, $\xi$ takes values between 1 and 9 at the end of inflation.~Thus exponential particle production occurs when the wavelength of the dark vector is comparable to the size of the horizon.~This generates a classical background dark electric field with a coherence length of order the inverse Hubble scale.~This dark electric field can then source Schwinger pair production during inflation.

To estimate how large the dark electric field is we can compute the energy density contained in the electric field.~Since the parameter $\xi$ increases with time, as $\dot\phi$ increases and $H$ decreases, it is largest at the end of inflation and in principle depends on time.~In~\cite{Bastero-Gil:2021wsf} the effect of this time dependence was taken into account in numerical solutions of~\eref{EOMfin} and shown that, while the time dependance can affect the shape and peak of the energy density power spectrum, the \emph{average} energy density is sufficiently well approximated using analytic solutions to obtain order of magnitude estimates of the dark matter parameter space.~Using the WKB approximation to analytically solve~\eref{EOMfin} one finds~\cite{Bastero-Gil:2018uel,Bastero-Gil:2021wsf},
\bea	\label{eq:rhoA}
\rho_E &\equiv&  \langle\rho_E\rangle 
= \frac{1}{2}\langle \frac{1}{a^2}|\partial_t A_+|^2\rangle \nn
&=& \frac{1}{2}\langle |E_+|^2\rangle 
\approx
10^{-4} \frac{e^{2\pi\xi}}{\xi^3} H^4 .
\eea
Here $\rho_E \equiv \langle \rho_E \rangle$ represents the spacial average of the total dark electric field energy density.~An accompanying dark magnetic field is also produced, but it is suppressed by $\rho_B = \rho_E/\xi^2$.~This magnetic field is automatically produced parallel to the electric field (see~\eref{polvecs} in the Appendix) and would thus lead to a small enhancement of the Scwhinger effect~\cite{Fujita:2022fwc}, but for the rough estimates of the parameter space we make here it can be neglected.

As shown in \cite{Bastero-Gil:2021wsf}, the analytic result in~\eref{rhoA} is in decent agreement with the numerical result which takes into account the time variation of $\xi$ with time.~Taking $\xi$ and $H$ constant, we see from~\eref{rhoA} that the magnitude of the physical dark electric field sourced by the rolling inflaton remains constant.~This is a reasonable approximation when we are close to the end of inflation as $\xi$ and Hubble change little in the last few e-folds.~Since the energy density is exponentially sensitive to $\xi$ which increases towards the end of inflation, as the slow roll parameter increases, the vast majority of the dark electric field is produced at the end of inflation.~Hence, in the last few e-folds of inflation we obtain a roughly constant dark electric field of order the Hubble horizon and we can apply~\eref{rhochi} and~\eref{rhochiapprox} to compute the energy density of the Schwinger produced dark charged particles.
 
To produce particles during inflation it is in general necessary to break conformal invariance.~We emphasize that here we consider a single source of conformal breaking generated by the axion like coupling between the inflaton and the dark photon, $\alpha/f$ in~\eref{Lsource} which introduces a new scale.~As discussed above, we take the dark charged particles $\chi$ to be conformally coupled by neglecting their mass during inflation, $m_\chi \ll H$ (and by taking $\xi_R = 1/6$ in~\eref{lagscal} for the scalar case).~This ensures there is no purely gravitational particle production~\cite{Ford:1986sy,Kuzmin:1998kk} and we can isolate the particle production due to the inflationary Schwinger effect.

\subsection{Conductivity $\&$ weak/strong electric force}\label{sec:sigma}

So far we have used the conductivity $\sigma$ to parametrize and quantify the energy density of the particles produced via the Schwinger effect.~Such an effect in flat spacetime depends on the  ``electric force''  ($g_D E$) as well as on the mass of the charged particle and leads to an exponential suppression $\propto{\rm exp}[-m_\chi^2/|g_D E|]$.~In de-Sitter space expansion introduces one more parameter in the Hubble scale $H$ and non exponentially suppressed regimes become possible.~In this section we give a brief account of how the conductivity is expected to behave in the $|g_D E| \ll H^2 $ (weak) and $g_D E \gg H^2$ (strong) electric force regimes.~For this we rely on previous studies of the Schwinger effect in a de-Sitter background~\cite{Kobayashi:2014zza,Hayashinaka:2016qqn,Banyeres:2018aax}.

The conductivity can be computed in terms of the two dimensionless ratios~\cite{Kobayashi:2014zza,Hayashinaka:2016qqn,Banyeres:2018aax},
\bea\label{eq:ratios}
\lambda = \frac{ g_D E }{ H^2 } , \quad \mbchi = \frac{ m_\chi }{ H} ,
\eea
with which one can study different limits.~The strong electric force regime is defined via the limit $\lambda \gg 1$.~In this regime the Schwinger effect exhibits the same behaviour in de-Sitter space as in flat space and leads to the dimensionless conductivity~\cite{Kobayashi:2014zza,Banyeres:2018aax},
\beq \label{eq:sigbarstrong}
\bar\sigma \equiv \frac{\sigma}{H} = \frac{g_D^2}{12\pi^3} \lambda e^{- \pi \mbchi^2/\lambda}
~~(\lambda \gg 1) .
\eeq
This applies to scalars, while for fermions the conductivity is larger by a factor of 2~\cite{Hayashinaka:2016qqn}.~When $\lambda > \mbchi$ we see the conductivity is not exponentially suppressed even if $m_\chi > H$.~This implies that particle production can occur during inflation via the Schwinger effect even if purely gravitational effects, which are exponentially suppressed with $m_\chi/H$, are negligible.~This case was studied in~\cite{Bastero-Gil:2023htv} which examined inflationary Schwinger production of superheavy dark matter with mass $m_\chi > \Hend$ to which we refer the interested reader.

Here we are interested in light dark matter where $\mbchi < 1 < \lambda$.~In the $\mbchi \gg \lambda$ regime $\sigb$ simplifies to,
\bea\label{eq:sigbarstrongmchi0}
\bar\sigma = \frac{g_D^2}{12\pi^3} \lambda ~~(\lambda \gg 1 \gg \mbchi) .
\eea
This implies that even in the massless (conformal) limit, where again purely gravitational effects are neglibile~\cite{Ford:1986sy,Kuzmin:1998kk} we can have particle production during inflation due to the Schwinger effect.~As we'll see below, this opens the possibility of generating light scalar and/or fermion dark matter during inflation which is typically difficult with only purely gravitational production.~Since we will only consider $\mbchi < 1$ in this study,~\eref{sigbarstrongmchi0} is a sufficient approximation for present purposes. 

In the weak force regime where $\lambda < 1$, no Schwinger production occurs in flat space while in de-Sitter space it can occur and exhibits various non-intuitive features~\cite{Hayashinaka:2016qqn,Banyeres:2018aax}.~Here we examine whether dark Schwinger production can also generate enough dark matter in the weak electric force regime.~In the weak electric force regime the conductivity is given by~\cite{Kobayashi:2014zza,Banyeres:2018aax},
\bea \label{eq:sigmascaling}
\bar\sigma = \frac{\sigma}{H} &=&  \frac{g_D^2}{24\pi^2} \left[ \ln 2 + \frac{2}{3} + 2 \gamma_E \right] \nonumber \\ 
&\approx& 10^{-2} g_D^2 \, ~~(\lambda \ll 1 ) ,
\eea
where, in order to ensure (purely) gravitational effects are negligible~\cite{Ford:1986sy,Kuzmin:1998kk}, we have set $m_\chi = 0$ and $\xi_R = 1/6$ in~\eref{lagscal} for a conformally coupled scalar and $R =12 H^2$ in de-Sitter.~The last equality holds approximately for both dark charged fermions and scalars~\cite{Kobayashi:2014zza,Hayashinaka:2016qqn,Banyeres:2018aax} and will be useful when exploring the dark matter relic density parameter space as a function of $g_D$.

\section{Cosmic evolution and dark\\ sector relic abundance}\label{sec:evolution}

At the end of inflation the inflaton still dominates the energy density of the Universe with,
\bea \label{eq:rhoIend}
\rhoIend = 3 \Hend^2 \Mpl^2 .
\eea
A small part of this initial energy density is transferred to the dark photon.~As discussed above, this energy is mostly stored in the classical dark electric field (see~\eref{rhoA}) with energy density at the end of inflation,
\bea \label{eq:rhoAend}
\rhoAend = \frac{1}{2} E^2
\approx 10^{-4} \frac{e^{2\pi \xiend}}{ \xiend^3 } \, \Hend^4 \, .
\eea
A small fraction of $\rhoAend$ then goes into pair producing $\chi$ via the dark Schwinger mechanism,
\bea\label{eq:rhochiend}
\rhochiend = \frac{2\sigend}{n}
\, \rhoAend .
\eea
The fraction of $\rhoIend$ which goes into visible radiation to 
reheat the universe, a process that we consider instantaneous here for simplicity, is parametrized by the dimensionless quantity $\epsilon_R$ defined by,
\bea \label{eq:rhoRH}
\rho_R (T_{\rm RH})  
= \frac{\pi^2}{30}g_*(\TRH) \TRH^4 \, 
\equiv \epsilon_R^4 \rhoIend  
= 3 \HRH^2 \Mpl^2.~~~~
\eea
Here $g_*(\TRH)$ denotes the number of relativistic degrees of freedom which we fix to $g_*(\TRH) \sim 100$, restricting ourselves to reheating temperatures above the electroweak scale.~We see~\eref{rhoRH} also defines the relation between the value of Hubble at reheating $H_{\rm RH}$ and $\Hend$ at the end of inflation,
\bea
H_{\rm RH} = \epsilon_R^2 \Hend \, ,
\eea
as well as the reheating temperature $\TRH$.~At reheating the dark vector and $\chi$ have energy densities,
\bea
\label{eq:rhoRHend}
\rho_A(\TRH) &=& \rho_E^{\rm end} \, , \\
\rho_\chi(\TRH) &=& \rho_\chi^{\rm end} \, ,
\eea
with the following hierarchies,
\bea \label{eq:hierarchy}
\rhochiend \ll \rhoAend \ll \rho_R(\TRH) \ll \rhoIend \, .
\eea
We assume that the dark and visible sectors are decoupled and that the dark sector does not thermalize with itself.~While at reheating $\rho_\chi$ starts as the smallest energy density, once the momentum of $\chi$ redshifts below its mass $\rho_\chi$ will begin to redshift like matter as $a^{-3}$ and eventually becomes dominant after matter radiation equality.

For the mechanism considered here, the dark vector could in principle be massless, in which case its energy density continues redshifting like radiation and remains negligibly subdominant throughout its cosmic evolution.~However, if the dark vector is massless limits from structure formation and plasma instabilities~\cite{Lasenby:2020rlf}, which are stronger than those coming from triaxiality of dark matter hallows~\cite{Lasenby:2020rlf,Agrawal:2016quu}, put severe constraints on the dark gauge couplings and would rule out much of the dark matter parameter space for inflationary Schwinger production.~These constraints can be avoided if the dark vector has a mass larger than $m_A \gtrsim 10^{-10}\,eV$ so at some point during its cosmic evolution, it must obtain a mass.~Thus in what follows, we consider a massive dark vector, in which case it eventually becomes non-relativistic and contributes to the dark matter relic abundance although for the masses we consider this contribution will be negligible.~The massless case is recovered by setting $m_A = 0$.

Let us now track the evolution of the energy densities, starting with the vector.~As shown in~\cite{Bastero-Gil:2018uel,Bastero-Gil:2021wsf}, for a given scale factor, the power spectrum of the dark vector will be peaked around scales the size of the horizon.~Thus, the modes which give the largest contribution to $\rho_A$ at reheating have physical momentum,
\bea\label{eq:qRH}
q_A(\TRH) \equiv  \frac{k}{\aend} \simeq \Hend \gg m_A\, .
\eea
As the universe expands and the temperature decreases the momentum redshifts as,
\bea\label{eq:qT}
q_A(T) = \frac{T}{\TRH} \Hend\, .
\eea
The dark photons then become non-relativistic when $q_A(\overline T_A) = m_A$ which occurs at the temperature,
\bea
\overline T_A 
= \frac{m_A}{\Hend} \, \TRH .
\label{eq:Tbar}
\eea
Above $\overline{T}_A$ their energy density redshifts like radiation, 
\bea \label{eq:rhoDrad}
\rho_A(T) = \rho_A(\TRH) \left( \frac{T}{\TRH} \right)^4 \, ,
\eea
while below $\overline T_A$ it redshifts like matter giving,
\bea \label{eq:rhoDmat}
\rho_A(T) = \rho_A(T_0) \left( \frac{T}{T_0}  \right)^3 \, ,
\eea
where $T_0 \approx 10^{-13}$\,GeV is today's CMB temperature.~From this we obtain the energy density today~\cite{Bastero-Gil:2018uel,Bastero-Gil:2021wsf},
\bea\label{eq:rhoA0}
\rho_A(T_0) = \left(\frac{T_0}{T_{RH}}\right)^3 \left(\frac{m_A}{\Hend} \right) \rhoAend   \, .
\eea
where we have used~\eref{rhoRHend}.~As shown in~\cite{Bastero-Gil:2018uel,Bastero-Gil:2021wsf}, the dark vector can account for the entirety of the dark matter over a very large range of dark vector masses $\sim\mu$eV $- 10$  TeV.~Furthermore, due to large density fluctuations, today the vector dark matter has a clumpy like nature on physical scales $\sim$ cm $ - 100$  km~\cite{Bastero-Gil:2021wsf}.~In what follows we will always consider $m_A \ll m_\chi < \Hend$ so that the dark photon contributes negligibly to the dark matter relic abundance which is dominated by $\chi$.

Turning now to the evolution of $\rho_\chi$ we must first determine the momentum of $\chi$ at the end of inflation (and instantaneous reheating) as done for the dark vector in~\eref{qRH}.~For the dark electrons which are Schwinger produced during inflation there are competing effects between damping due to expansion and the acceleration due to the background dark electric field.~We can write the evolution of the physical momentum of $\chi$ as,
\bea
\label{eq:qinf}
q(\tau)_\chi = q(\tau_0)_\chi\,e^{-N} + \lambda H (1 - e^{-N}) ,
\eea
where $q(\tau_0)$ is the momentum at production in conformal time which depends on the electric field production mechanism and $N$ counts the number e-folds from $\tau_0$.~The first term is the usual damping term due to expansion.~The second term is due to the presence of the background electric field which accelerates the electrons.~The net effect is that during inflation $\chi$ very quickly reaches a terminal momentum of $\lambda H$.~This holds generically for any mechanism which produces an electric field during inflation that sources Schwinger production.

However, for the dark electric field production mechanism considered above and examined in~\cite{Bastero-Gil:2021wsf}, the dark electric field is dominantly produced just at the end of inflation.~In this case the dark electrons have not had much time to either redshift or get accelerated by the electric field.~Thus taking $N\approx 0$ and assuming the dark electrons are produced with momentum comparable to the dark photons, at the end of inflation (and instantaneous reheating) we again have for the initial momentum,
\bea\label{eq:qRH}
q_\chi(\TRH) \simeq \Hend \gg m_\chi \, .
\eea
A precise determination of $q_\chi(\TRH) $ requires numerical study, but could potentially further expand the dark matter parameter space examined below.

As we detail in the Appendix, once inflation ends the electric field quickly begins to oscillate~\cite{Bastero-Gil:2021wsf} and no longer accelerates the dark electrons efficiently.~Thus as the universe expands and the temperature decreases the momentum for the dark electrons redshifts as in~\eref{qT} for the dark photon, but with initial momentum given in~\eref{qRH}.~Thus for the energy density of $\chi$ today we can follow the same steps as for the evolution of the dark photon and immediately write,
\bea\label{eq:rhochi0}
\rho_\chi(T_0) = \left(\frac{T_0}{T_{RH}}\right)^3 \left(\frac{m_\chi}{\Hend} \right) \frac{\sigb}{2} \rhoAend  \, ,
\eea
where we have used~\eref{rhochiend} and~\eref{rhoRHend} for $\rhochiend$.~This analysis is sensible only if the dark sector does not thermalize with the visible sector.~Thermalization with the visible sector is easily avoided by taking the dark sector to be secluded.~This implies that a possible kinetic mixing between visible and dark $U(1)$ has to be small.

\subsection{Constraints on dark gauge coupling}\label{sec:gDconstraints}

The above analysis assumes the different dark sector components redshift independently, thus we also need to ensure that the dark sector does not thermalize within itself.~This requires that the rate of interactions between the dark charged particles and the dark photon is always slower than Hubble which we can estimate as follows.~While $\chi$ is relativistic its number density is given by $n_\chi(T) \simeq \rho_\chi(T) / q_\chi(T)$ and the 2-to-2 cross section, mediated by a dark photon, can be estimated as $\sigma_{2\to2} \sim g_D^4/ q^2_\chi(T)$.~The scattering rate is $\Gamma = n_\chi(T) \sigma_{2\to2}$.~Requiring that $\Gamma(T) < H(T)$ until $\chi$ becomes non-relativistic at a temperature $\Tbchi = \frac{m_\chi}{\Hend} \TRH$ leads to the condition $g_D^4 \sigb < \epsilon_R^2 \mbchi (\Hend^2/E)^2$ where $\sigb$ is the conductivity given in~\eref{sigbarstrong} and~\eref{sigmascaling} for the $\lambda > 1$ and $\lambda < 1$ regimes respectively.

In the strong electric force limit $\lambda > 1$ the no thermalization in the dark sector condition leads to,
\bea
\label{eq:gDTHERMstrongR}
g_D \lesssim 2\left( \frac{ \mbchi \epsilon_R^2}{(E/\Hend^2)^3} \right)^{1/7} ~~ (\lambda > 1).
\eea
An additional upper bound on $g_D$ comes from requiring no backreaction from Schwinger production which could destroy the electric field and which implies $\sigb < 1$, but this leads to a weaker bound than~\eref{gDTHERMstrongR}.~A lower bound on $g_D$ comes from imposing $\lambda > 1$,
\bea
g_D > \frac{1}{E/\Hend^{2}} ~~ (\lambda > 1).
\eea
In the weak electric force limit the strongest upper bound comes from imposing the condition $\lambda < 1$,
\bea
g_D < \frac{1}{E/\Hend^{2}} .
\eea
An additional lower bound in both cases arises if we demand we satisfy the weak gravity conjecture~\cite{Arkani-Hamed:2006emk}, 
\bea\label{eq:WGC}
g_D > \frac{m_\chi}{M_{\rm Pl}} \, .
\eea

As discussed above, in principle there are severe constraints on $g_D$ coming from structure formation and plasma insatiabilities~\cite{Lasenby:2020rlf}, but these can be avoided if the dark vector has a mass $m_A \gtrsim 10^{-10}\,eV$.~Assisted Schwinger production~\cite{Siemonsen:2022ivj} due to accelerated dark electrons could potentially destroy the dark electric field and constrain $g_D$, but since the dark electrons reach terminal velocity (see~\eref{qinf}) so quickly during inflation and are not accelerated afterwards (see Appendix) we expect this to also have little effect.~When examining the dark matter parameter space below we thus only impose on $g_D$ the constraints in~\eref{gDTHERMstrongR}-\eref{WGC}.

\subsection{Dark matter parameter space}\label{sec:pspace}

With the various energy densities today in hand we can examine the parameter space in which the observed dark matter relic abundance is reproduced.~Taking the observed energy density of cold dark matter today~\cite{Ade:2015xua} to be $\rho_{\rm CDM} = 9.6 \times 10^{-48} \ {\rm GeV}^4$, the relic abundances of the dark vector and dark electrons are given by,
\bea\label{eq:relic}
\frac{\Omega_A}{\Omega_{\text{CDM}}} &=&
\frac{\rho_A(T_0)}{\rho_{\text{CDM}}}   , \quad
 \frac{\Omega_\chi}{\Omega_{\text{CDM}}} =
 \frac{\rho_\chi(T_0)}{\rho_{\text{CDM}}} \, , 
\eea
where  $\rho_A(T_0)$ and $\rho_\chi(T_0)$ are given in~\eref{rhoA0} and~\eref{rhochi0} respectively.~With these estimates for the relic abundance, we can go on to obtain the regions of parameter space for a viable dark matter sector.~In addition to the constraints on $g_D$ discussed above, there are additional constraints on the model parameters which must be satisfied for consistency of the mechanism.

The first is that we must also require the hierarchy in energy densities at reheating in~\eref{hierarchy}.~Imposing that the visible radiation has energy density lower than the inflaton at the end of inflation ($\rho_R \ll \rhoIend$) imposes $\epsilon_R < 1$.~To ensure the Universe is radiation dominated until matter radiation equality requires that $\rho_R > \rhoAend$ and translates into an upper bound on
$\Hend$,
\bea\label{eq:Hendbound}
\Hend < \frac{ \sqrt{6} \, \epsilon_R^2  }{(E/\Hend^2)} M_{\rm Pl}  \, .
\eea
So we see that for very large electric fields or reheating temperatures which are low $(\epsilon_R \ll 1)$, the upper bound on the Hubble scale at the end of inflation becomes very stringent.~We impose the stronger between this upper bound and the absolute astrophysical upper bound of $\Hend \lesssim 10^{13}$\,GeV which must be smaller than the inflationary scale early during inflation.~We must also ensure no backreaction on the inflaton dynamics from production of the dark electric field.~This requires $\rhoAend < \rhoIend$ which restricts the size of the electric field to~\cite{vonEckardstein:2023gwk},
\bea\label{eq:ratiobound}
1 < \frac{E}{\Hend^2} < 10^4 .
\eea
Note that for the vector production mechanism discussed above, this corresponds to $1.6 \lesssim \xiend \lesssim 5$.~The dark electron mass is constrained from below by requiring that $\chi$ becomes non-relativistic before matter radiation equality, thus behaving like cold dark matter.~In practice the no-thermalization condition imposes stronger lower bounds on $m_\chi$.~There are also robust constraints coming from astrophysical probes which in the case of fermions can lead to a stronger and absolute lower bound of $\mathcal{O}(100 \ {\rm eV})$~\cite{DiPaolo:2017geq,Alvey:2020xsk} for fermion dark matter.

We focus on the light mass case where $m_\chi < \Hend$ while the large mass case $m_\chi > \Hend$ was examined in~\cite{Bastero-Gil:2023htv} to which we refer the interested reader.~We assume the dark photon mass is larger than $m_A \gtrsim 10^{-10}$\,eV to avoid severe constraints on $g_D$ coming from plasma instabilities~\cite{Lasenby:2020rlf}, but small enough that it contributes negligibly to the dark matter relic abundance.

\subsubsection{Weak dark electric force $\lambda < 1$}

We first consider the weak electric force regime where $\lambda < 1$.~Note this case is not possible in Minkowski space which requires a strong electric force and is a unique feature in de-sitter space~\cite{Banyeres:2018aax}.~In this regime the upper bound on $g_D$ coming from imposing $\lambda < 1$ is a stronger constraint than both the no thermalization and no Schwinger backreaction conditions.~In~\fref{DMplots1} we show contours of $\Omega_\chi/\Omega_{\rm{CDM}} = 1$ given in~\eref{relic} as a function of $m_\chi$\,vs.\,$\Hend$ for various values of $E/\Hend^2$ and $\epsilon_R$ as indicated.~For the conductivity in the $\lambda << 1$ regime we take the conformal limit $m_\chi = 0$ (and $\xi_R = 1/6$ in the scalar case defined in~\eref{lagscal}) leading to~\eref{sigmascaling}.~Note since we are in the conformal limit, purely gravitational production is negligible~\cite{Ford:1986sy,Kuzmin:1998kk}. 
%
\begin{figure*}[tbh]
\begin{center}
\includegraphics[scale=0.34]{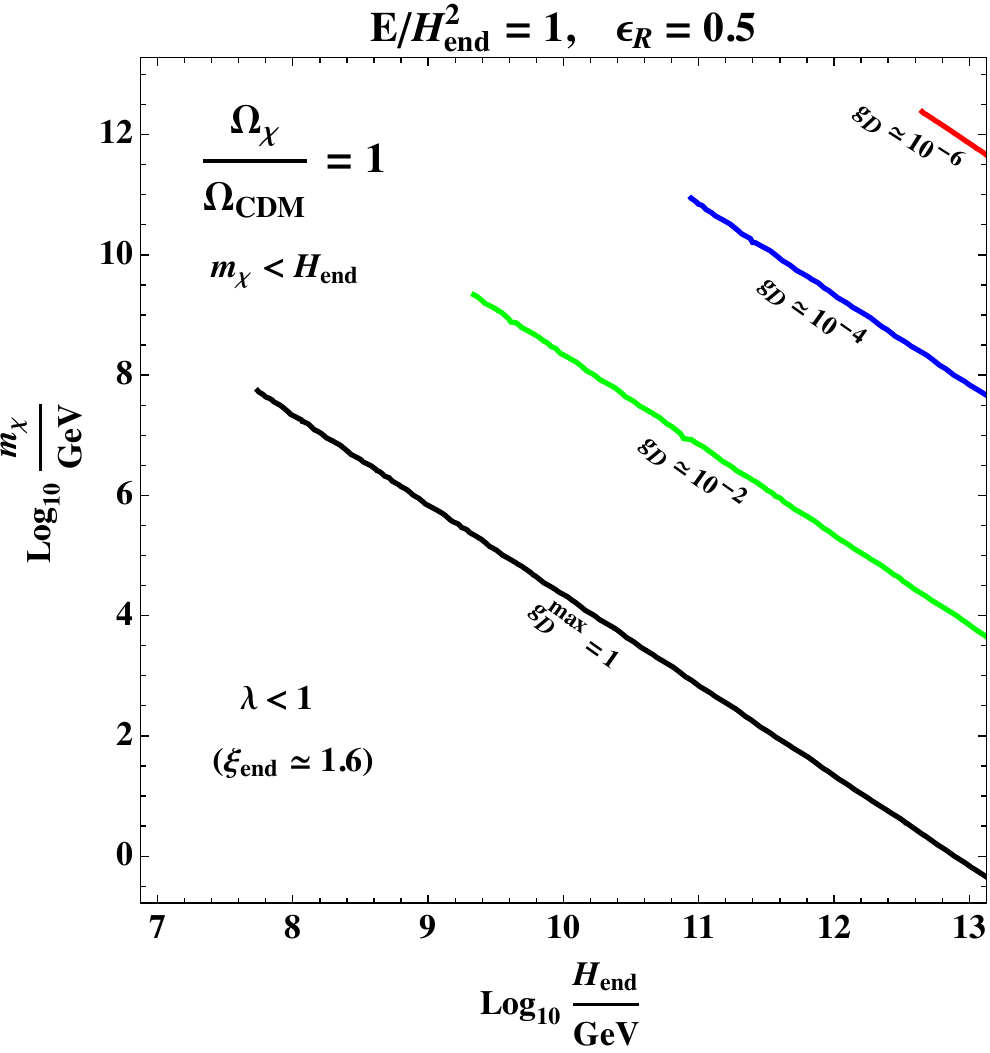}
\includegraphics[scale=0.34]{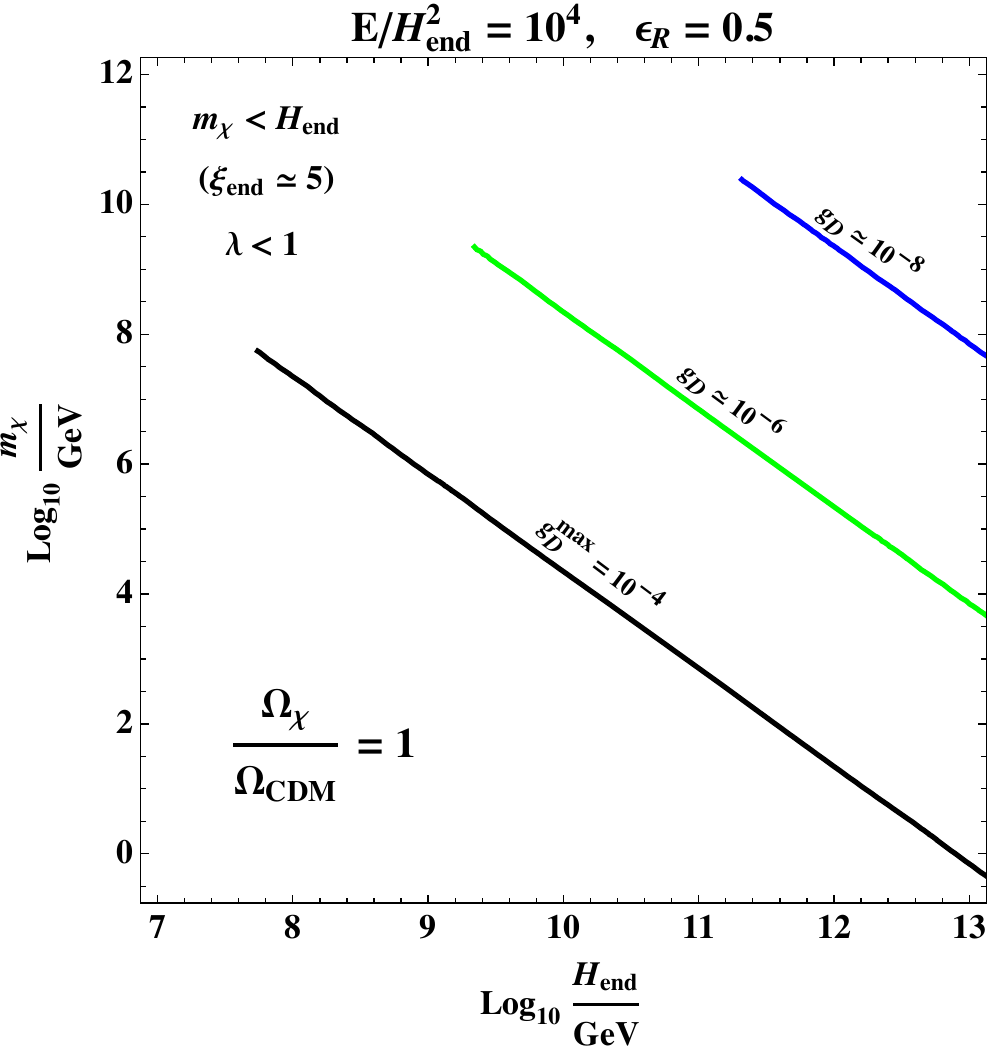}
\includegraphics[scale=0.33]{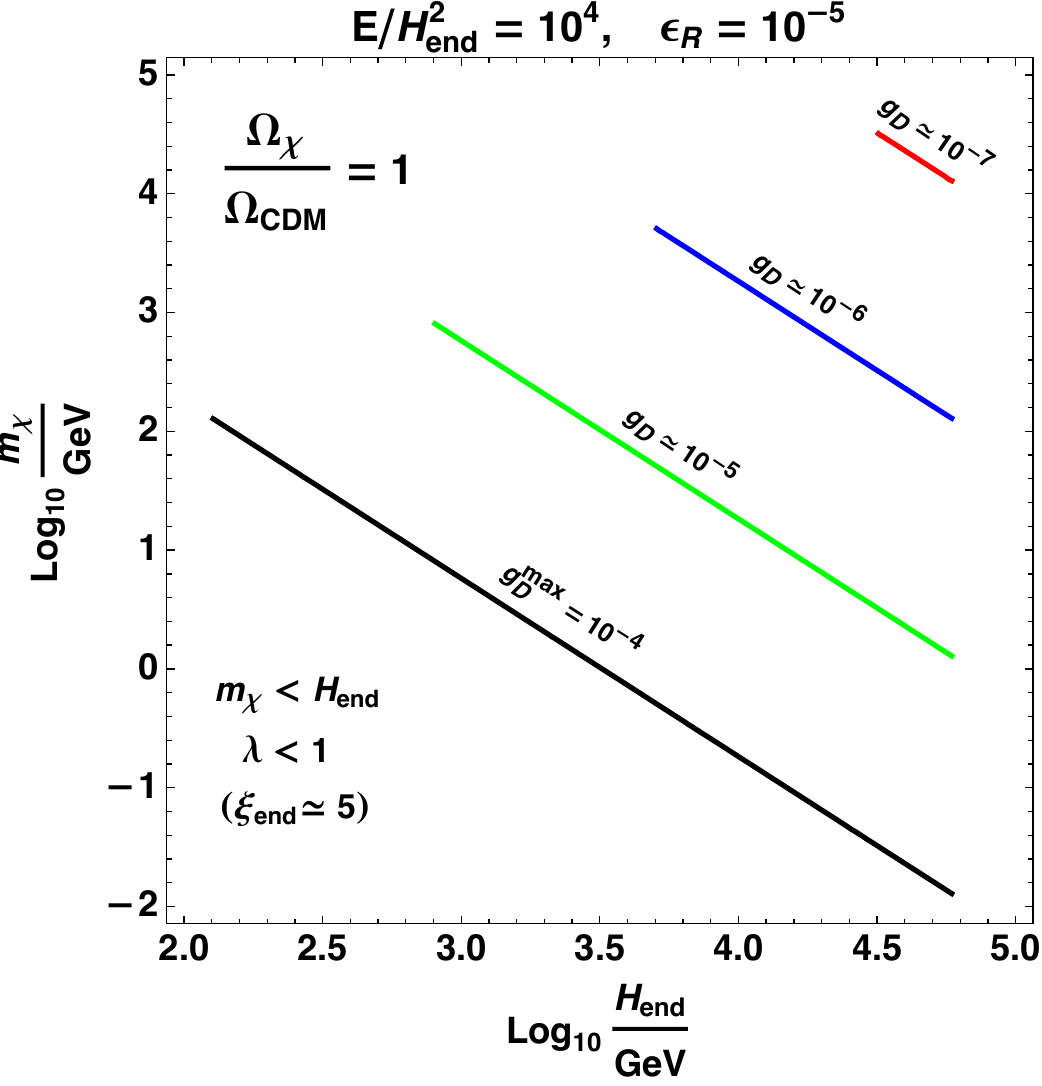}
\end{center}
\caption{Contours of $\Omega_\chi/\Omega_{\rm{CDM}} = 1$ given in~\eref{relic} as a function of $m_\chi$\,vs.\,$\Hend$ in the weak electric force regime where $\lambda < 1$ for various values of $E/\Hend^2$, $\epsilon_R$, and $g_D$ as indicated.~See text for more information.
}
\label{fig:DMplots1}
\end{figure*}
%

In the left plot we show contours for $E/\Hend^2 = 1$ and $\epsilon_R = 0.5$.~For the dark electric field production mechanism discussed above, $E/\Hend^2 = 1$ corresponds to $\xiend \simeq 1.6$ (see~\eref{rhoAend}).~Note that for $\epsilon_R = 0.5$ this range of Hubble scales corresponds to reheating temperatures between $\TRH \sim (10^{9} - 10^{15})$\,GeV (see~\eref{rhoRH}).~For $E/\Hend^2 = 1$ the maximum value of $g_D$ allowed by imposing $\lambda < 1$ is given by $g_D^{\rm{max}} =1$ (black) for the which the dark matter relic abundance can be reproduced over a very large range of masses from $m_\chi \sim 100$\,MeV to $m_\chi \sim 10^{8}$\,GeV.~For this parameter point, dark gauge couplings as small as $g_D \simeq 10^{-6}$ can reproduce the dark matter relic abundance for dark matter masses $m_\chi \sim 10^{12}$\,GeV and $\Hend \sim 10^{13}$\,GeV.~Much smaller values of $g_D$ would require $\Hend \gtrsim 10^{13}$\,GeV which is excluded by astrophysical measurements. 

In the middle plot we show contours for $E/\Hend^2 = 10^4$  ($\xiend \simeq 5$) with $\epsilon = 0.5$.~Now the maximum value of $g_D$ allowed by imposing $\lambda < 1$ is $g_D^{\rm{max}} =10^{-4}$ (black).~Overall the parameter space is similar to the top left figure with $E/\Hend^2 = 1$ and $\epsilon = 0.5$ except now we are pushed to smaller gauge couplings which can be as small as $g_D \simeq 10^{-9}$ for $m_\chi \sim 10^{10}\,\Gev$.  

On the right we show contours for $E/\Hend^2 = 10^4$ with $\epsilon_R = 10^{-5}$.~In this case we have lower reheating temperatures in the range $\TRH \sim (10^{4} - 10^{6})$\,GeV.~This forces us to smaller Hubble scales in order to ensure $\rhoAend < \rho_R$.~This in turn lowers $m_\chi$ in order to ensure $m_\chi < \Hend$.~We see that for dark gauge couplings in the range $g_D \simeq (10^{-4} - 10^{-7})$ the relic abundance can be reproduced for dark matter masses in the range $m_\chi \sim 10\,\Mev - 100$\,TeV and Hubble scales at the end of inflation in the range $\Hend \simeq (100 - 10^{5})$\,GeV.

To summarize, in the weak force regime with $\lambda < 1$, inflationary dark Schwinger production can reproduce the dark matter relic abundance for dark electron masses in the range $m_\chi \simeq 10\,\Mev - 10^{12}$\,GeV and gauge couplings between $g_D \simeq 10^{-9} - 1$ for Hubble scales at the end of inflation in the range $\Hend \simeq (100 - 10^{13})$\,GeV.

\subsubsection{Strong dark electric force $\lambda > 1$}

In~\fref{DMplots2} we show contours of $\Omega_\chi/\Omega_{\rm{CDM}} = 1$ in the strong electric force regime where $\lambda > 1$.~In this regime the Schwinger conductivity is the same as for flat space~\cite{Banyeres:2018aax} and is the same for both fermions and scalars (up to a factor of 2).~The $\lambda > 1$ condition now imposes a lower bound on $g_D$ while an upper bound (blue shaded) comes from requiring that the dark sector does not thermalize with itself (see~\eref{gDTHERMstrongR}).~There is also an upper bound from requiring no Schwinger backreaction (red shaded), but it is weaker than requiring no thermalization.~In principle the parameter space between these two bounds could also be viable dark matter parameter space, but one would need to track the thermal history after inflation.~This involves other strong dark QED processes~\cite{Arvanitaki:2021qlj} in order to compute the proportion of dark electrons to photons in the final relic abundance and is beyond the scope of this work.~We show this region for illustrative purposes, but when considering the viable dark matter parameter space presented here we take the no-thermalization condition to give the upper bound on $g_D$. 
%
\begin{figure*}[tbh]
\begin{center}
\includegraphics[scale=0.34]{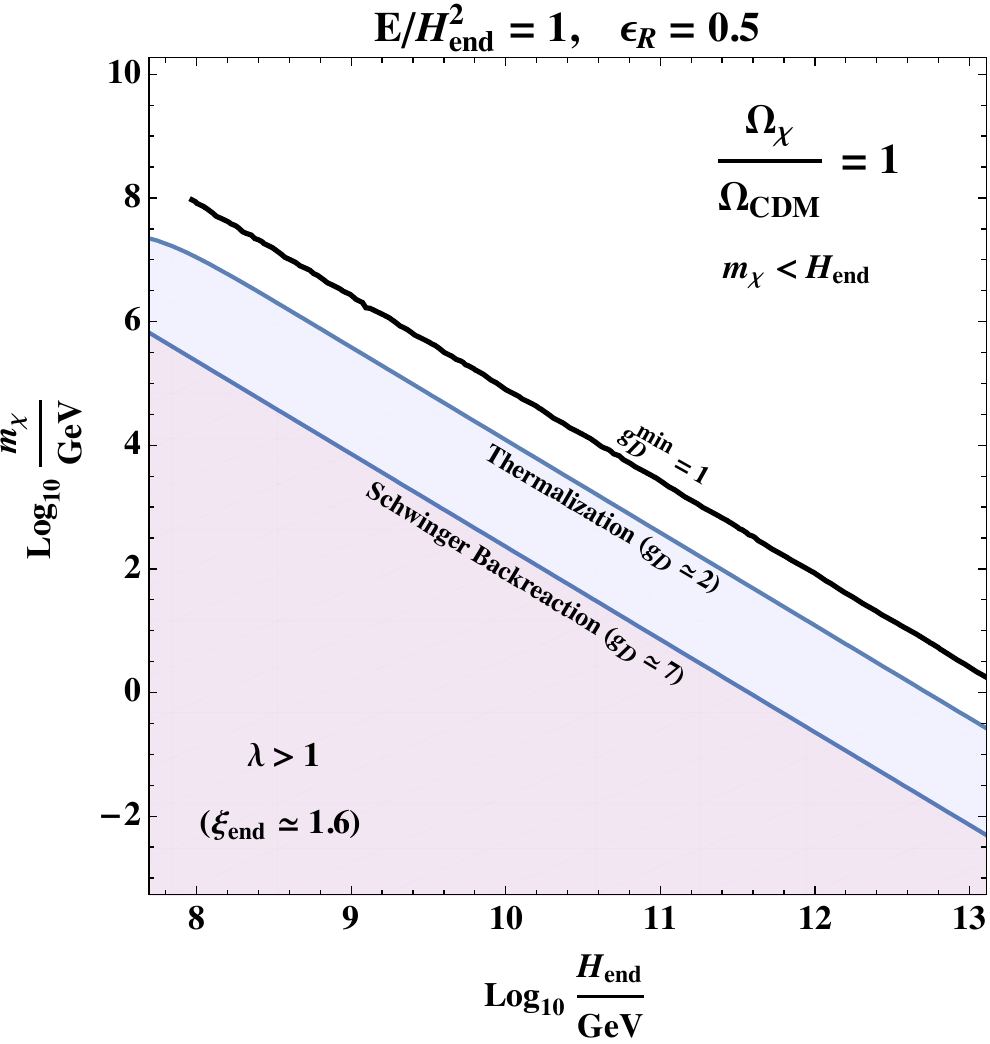}
\includegraphics[scale=0.35]{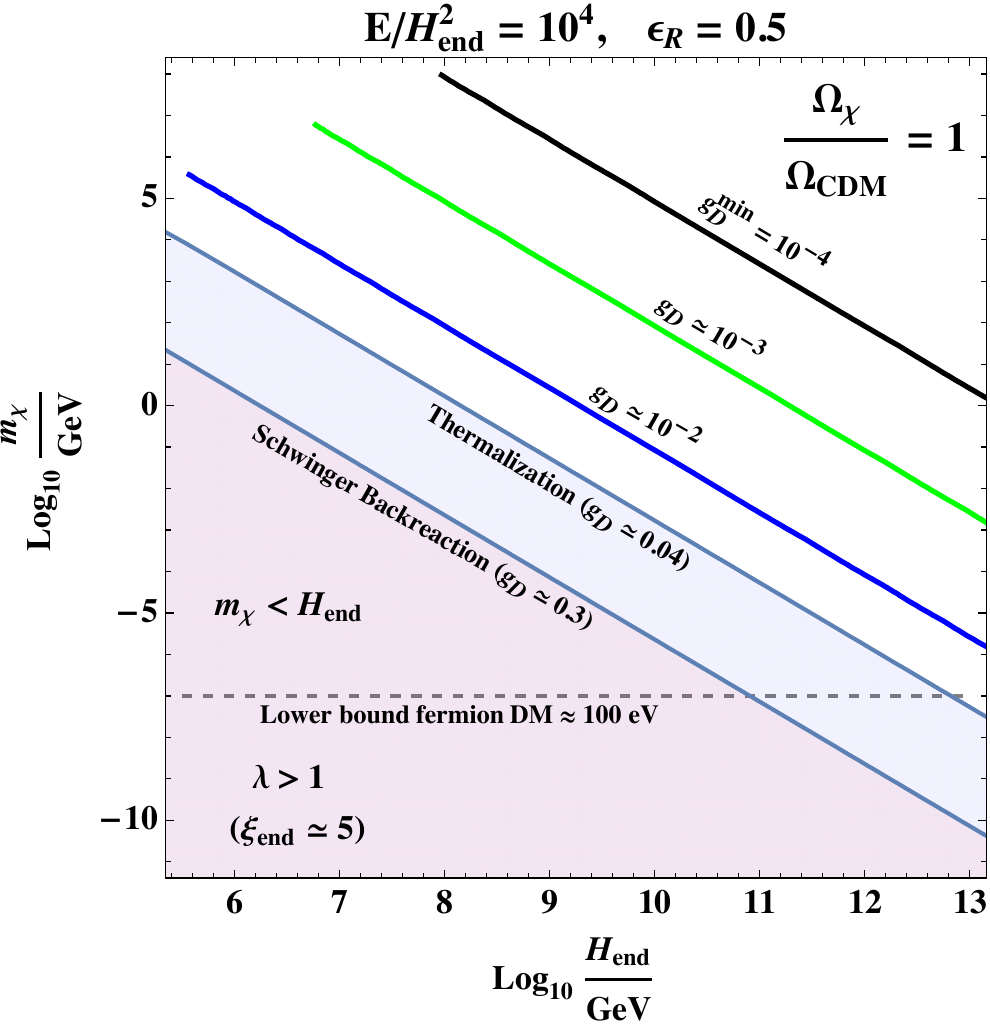}
\includegraphics[scale=0.35]{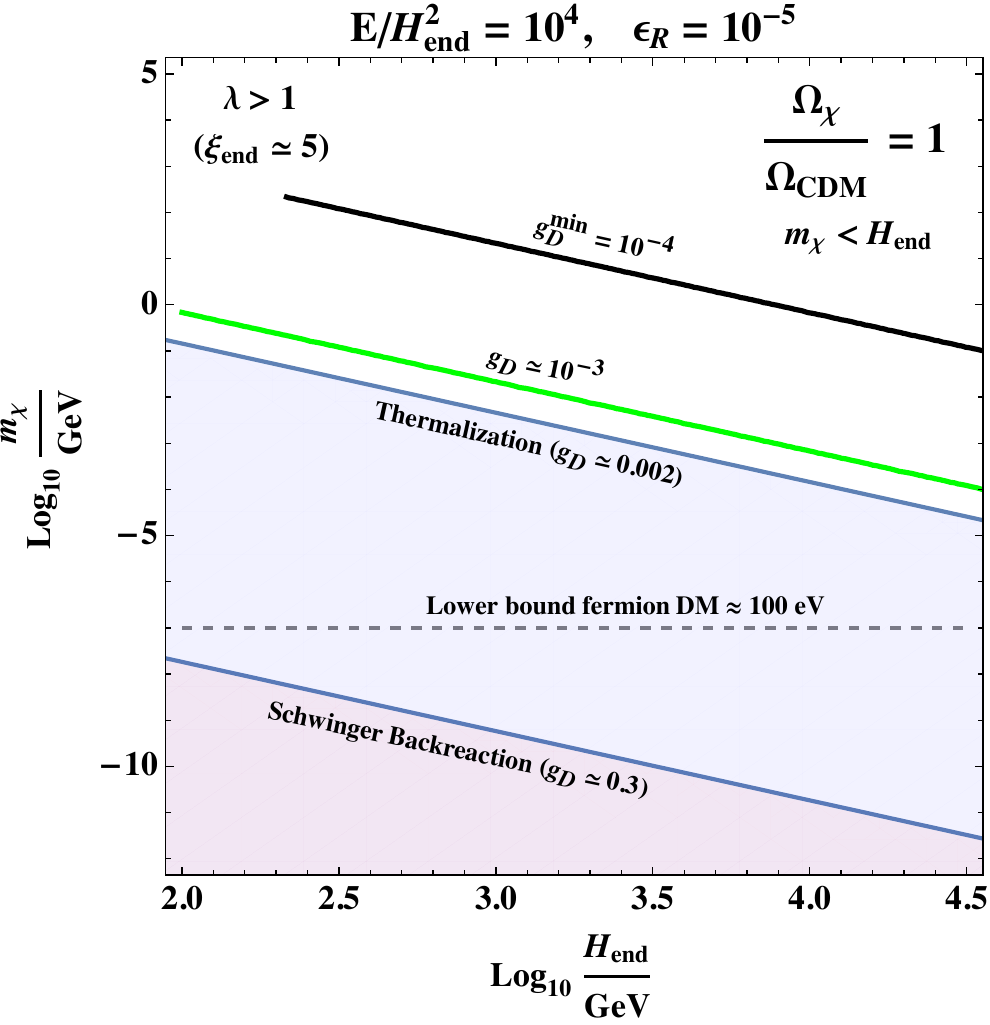}
\end{center}
\caption{Contours of $\Omega_\chi/\Omega_{\rm{CDM}} = 1$ given in~\eref{relic} as a function of $m_\chi$\,vs.\,$\Hend$ in the strong electric force regime where $\lambda > 1$ for various values of $E/\Hend^2$, $\epsilon_R$, and $g_D$ as indicated.~See text for more information.
}
\label{fig:DMplots2}
\end{figure*}
%

In the left plot we show contours again for $E/\Hend^2 = 1$ and $\epsilon_R = 0.5$.~We see in this case there is a narrow band between $g_D \simeq 1 - 2$ for which the relic abundance can be reproduced.~Along this band however the relic abundance can be reproduced over a large range of dark matter masses from $m_\chi \simeq (1 - 10^{8})$\,GeV for Hubble scales at the end of inflation $10^{8}\,\Gev \lesssim \Hend \lesssim 10^{13}\,\Gev$.
 
In the middle plot we show contours for $E/\Hend^2 = 10^4$ with $\epsilon = 0.5$.~We see dark electrons as light as $m_\chi \sim 100$\,eV can give the correct relic abundance, but now for dark gauge couplings near the thermalization bound of $g_D \simeq 0.04$ and Hubble scales $\Hend \sim 10^{13}$\,GeV.

Finally, on the right we show contours for $E/\Hend^2 = 10^4$ with $\epsilon_R = 10^{-5}$.~Again the lower reheating temperatures force us to smaller Hubble scales in order to ensure $\rhoAend < \rho_R$.~In the end the relic abundance can still be reproduced for a narrow band $g_D \simeq 10^{-4} - 10^{-3}$ and  $\Hend \simeq (100 - 10^{5})$\,GeV leading to a dark matter mass in the range $m_\chi \simeq 100\,\kev - 100$\,GeV.

In summary, in the $\lambda > 1$ strong force regime we can have dark charged scalars as the dark matter for masses as light as $m_\chi  \sim 0.1$\,eV and $m_\chi \sim  100$\,eV for fermions and as heavy as $m_\chi \sim 10^{12}$\,GeV in both cases.~The dark gauge coupling can span the range $g_ D \sim 10^{-4} - 1$ while the Hubble scale at the end of inflation can span the entire range allowed by astrophysical bounds.

\section{Summary and Discussion} \label{sec:sum} 

We have presented an inflationary Schwinger production mechanism which can generate dark matter during inflation even if purely gravitational production is negligible.~We have shown (see~\fref{DMplots1} and~\fref{DMplots2}) that the observed dark matter relic abundance can be generated via a \emph{dark} Schwinger mechanism based on a `dark QED' $U(1)$ gauge theory which takes place during inflation.~This mechanism leads to a feebly interacting dark sector comprised of dark charged fermions and/or scalars which can be as light as $m_\chi \sim 0.1$\,eV for scalars and $m_\chi \sim 100$\,eV for fermions and as heavy as $m_\chi \sim 10^{12}$\,GeV in both cases.~In the small mass limit we consider the dark photon which makes up the electric field contributes negligibly to the relic abundance, but must at some point obtain a mass $m_A \gtrsim 10^{-10}$\,eV in order to evade constraints from structure formation.~The dark gauge coupling can have a range $10^{-9} \lesssim g_D \lesssim 1 $ while the Hubble scale at the end of inflation can span $100\,{\rm GeV}  \lesssim \Hend \lesssim 10^{13}$\,GeV with reheating temperatures in a similar range.~Overall, we see a very large parameter space for which inflationary Schwinger production can accommodate the observed dark matter relic abundance.

The necessary background dark electric field is generated during inflation and sources Schwinger production until the end of inflation.~We have discussed possible sources for this dark electric field during inflation with particular focus on the mechanism introduced in~\cite{Bastero-Gil:2018uel,Bastero-Gil:2021wsf}.~We have derived the energy density of the Schwinger pair produced dark charged particles at the end of inflation (see~\eref{rhochiapprox}).~We then derived the relic abundance of the dark sector today (see~\eref{relic}) and examined contours where the amount of dark electrons matches the observed dark matter relic abundance.

We have emphasized that in inflationary Schwinger production the background electric field serves as a source of conformal symmetry breaking which is present even in the absence of conformal breaking in the fermion and scalar sectors.~Thus even massless fermions or conformally coupled massless scalars, which would not be produced by purely gravitational particle production, can still be produced during inflation via the Schwinger effect.~The dark matter has a power spectrum which is peaked at very small scales, thus evading potential isocurvature constraints from measurements of the CMB.~Our analysis relied on analytic calculations of the conductivity associated with the conserved Schwinger current during inflation computed in various studies of inflationary magnetogenesis~\cite{Hayashinaka:2016qqn,Banyeres:2018aax}.

The non-thermal dark QED sector contained in this mechanism could have a number of interesting phenomenological possibilities depending on how the dark sector communicates with the Standard Model, if at all.~For instance a coupling to the Standard Model could occur through kinetic mixing with the photon.~Another possibility is through Higgs mixing or through the neutrino sector.~Even if the dark sector does not communicate with the Standard Model it could have implications for intermediate scale structure~\cite{Archidiacono:2022iuu, Bottaro:2023wkd, Bottaro:2024pcb}.~Since the dark photon must obtain atleast a small mass there could be gravitational waves associated with a dark Higgs phase transitions in the dark sector.~Furthermore, we can have a multi-component dark sector, thermalized with itself or not, composed of fermions/scalars exchanging an ultralight vector which could have interesting effects on dark matter phenomenology.~All of these possibilities are potentially interesting and deserve further investigation.


~\\

\noindent
{\bf Acknowledgments:}~The authors thank Antonio Torres Manso, Manel Masip, Jose Santiago and Takeshi Kobayashi for useful comments and discussions.~The authors would also like to thank the referees who's comments led to much improved results and enlarged parameter space.~This work has been partially supported by Junta de Andaluc\'a Project A-FQM-472-UGR20 (fondos FEDER) (R.V.M.) and by SRA (10.13039/501100011033) and ERDF
under grant PID2022-139466NB-C21(R.V.M.) as well as by MICINN (PID2019-105943GB-I00/AEI/10.13039/501100011033,PID2022.140831NB.I00
/AEI/10.13039/501100011033/FEDER,UE)\,(M.B.G.),\,
FCT-CERN\,grant\,No.\,CERN/FIS-PAR/0027/2021 (M.B.G., P.F.),\,FCT\,Grant\,No.SFRH/BD/151475/2021 (P.F.).~The work of LU was supported by the Slovenian Research Agency under the research core funding No. P1-0035 and in part by the research grant J1-4389.~This article is based upon work from COST Action COSMIC WISPers CA21106, supported by COST (European Cooperation in Science and Technology).~RVM is grateful to the Mainz Institute for Theoretical Physics (MITP) of the Cluster of Excellence PRISMA$^+$ (Project ID 390831469), for its hospitality and partial support during the completion of this work.

\begin{section}{Appendix}

In this Appendix we study the dark electric field after inflation and determine how the dark electron momentum evolves.~We are interested in the tachyonic production of the dark photon discussed above (see~\eref{EOMfin}) in which only one transverse component is produced which we call $A_+$.~We work with the FLRW metric,
\bea
ds = dt^2 - a(t)^2 d \vec x^2 = a(\tau)^2 (d \tau^2 - d \vec x^2)\,,
\eea
where $a(t)$ is the scale factor, and $\tau$ the conformal time, $d \tau= dt/a(t)$.~The vector potential can be separated into transverse $\vec A_T$ and longitudinal $A_L$ components, such that  $\vec k \cdot \vec A= k A_L$ and  $\vec k \cdot \vec A_T= 0$.~Using the notation $\pm$ for the 2 transverse components, the (classical) vector field can be written as,
\bea
\vec A = \sum_{\lambda=\pm, L} \int \frac{d^3 k }{(2 \pi)^{3/2}} ( \vec e_\lambda A_\lambda (t) e^{i \vec k \cdot \vec x} + h. c. ) \,,
\eea
where $\vec e_\lambda$ are the longitudinal/transverse polarization unit vectors,
\bea
\vec k \cdot \vec e_L=k \,,  \;\;\; \vec k \cdot \vec e_{\pm} =0 \,,\;\;\; \vec k \times \vec e_{\pm}= \mp i k \vec e_{\pm} \,. \label{eq:polvecs}
\eea
Since the electric and magnetic fields are defined as,
\bea
a^2 \vec B = \vec \nabla \times \vec A \,, \;\;\; a^2 \vec E= - \partial_\tau \vec A= -a \partial_t \vec A \,,
\eea
the last condition in~\eref{polvecs} ensures that the (transverse) electric and magnetic fields produced during inflation are (anti)parallel.~Thus the presence of the magnetic field would only serve to enhance the Schwinger effect~\cite{Fujita:2022fwc}.

Taking for example the comoving momentum in the $z$ direction, $\vec k= (0,0,k)$, the longitudinal and transverse unit vectors can be chosen as,
\bea
\vec e_L&=&(0,0,1) \,,\\
\vec e_+ &=& \frac{1}{\sqrt{2}} (1, i, 0) \,,\\
\vec e_- &=& \frac{1}{\sqrt{2}} (i, -1, 0) \,.
\eea
The transverse potential vector components are then given by:
\bea
A_+^x &=& \frac{1}{\sqrt{2}}\int_k (A_+ e^{i \vec k \cdot \vec x} + h.c)\nonumber \\
&=& \frac{2}{\sqrt{2}} Re \int_k A_+ e^{i \veck \cdot \vec x}  
= \int_k A^R_+ \cos(k z) \,, \\
A_+^y &=& \frac{i}{\sqrt{2}}\int_k (A_+ e^{i \vec k \cdot \vec x} - h.c)= - \int_k A^I_+ \sin(k z) \,,
\eea
and similarly for their time derivatives.~From numerical simulations, once the perturbation re-enters the horizon ($k/a > H$) and the field starts oscillating we find that $A_+^R \sim \pm A_+^I$,  $\dot A_+^R \sim \pm \dot A_+^I$.

The spectrum of $A_+$ modes is peaked at the end of inflation around $k_P \simeq \mathcal{O}(a_{end} H_{end})$, so let  us approximate their amplitudes by a delta function
around the peak $k_P$ times an oscillatory function with frequency $k \tau$, 
\bea
A_+^x &=& A_+(0) \cos(\Omega_k) \cos(k z) \,, \\
A_+^y &=& \pm A_+(0) \cos ( \Omega_k)\sin(k z) \,,
\eea
$A_+(0)$ being the initial amplitude of the oscillations and,
\bea
\Omega_k \simeq k_P \tau = k_P \int \frac{dt}{a} = k_P \int \frac{da}{a^2 H}= \frac{k_P}{H_0} (\frac{a}{a_0} -1) \,,~~~~
\eea
where in the last equality we have assumed that the universe is radiation dominated.~The physical electric field, given by $\vec E = - \partial_\tau \vec A= -\partial_t \vec A/a$, would be given by,
\bea
E^x \simeq E^x_+(t) \sin(\Omega_k(t)) \cos(k z) \,, \\
E^y \simeq \pm E^y_+(t) \sin(\Omega_k(t)) \sin(k z) \,,
\eea
$E^i_+$ being the amplitude of the oscillations.~We have kept the superindex $i=x,\,y$ because the initial amplitudes may not be exactly the same.~We know that once inflation and the tachyonic resonance end, the energy density of the  dark vector  will behave as radiation and therefore\footnote{We use the subindex ``0'' to denote the values at the start of the oscillations.} $E^i_+ = E^i_0/a^2$.

The physical momentum of the charged particles follows the evolution equation,
\bea
\frac{d p^i}{d t} + H p^i \simeq g_D E^i \,,
\eea
where we have neglected the (subdominant) magnetic field contribution.~For example, for the ``x'' component, and dividing by $H$, we have,
\bea
\frac{d p^x}{d \ln a} + p^x = \frac{g_D E^x_0}{H_0} \sin(\Omega_k(a/a_0)) \cos(k z) \,,
\eea
where again we have used the fact that both $E_+(t)$ and $H(t)$ redshift as $a^{-2}$. Using $p^x=\bar p^x/a$ and $u=a/a_0$ we have:
\bea
\frac{d \bar p^x}{d u }= \frac{g_D E^x_0}{H_0} \sin(\frac{k_P}{H_0} (u -1)) \cos(k z) \,,
\eea
and then,
\bea
p^x &=& \frac{a_0}{a}\bar p^x \nonumber \\
&=& \frac{a_0}{a} \left[ p^x(0) + \frac{g_D E^x_0}{k_P} 2\sin^2(\frac{k_P}{H_0} (\frac{a}{a_0} -1))\cos(k z) \right] \,,~~~~~
\eea
where $p^x(0)$ is the initial value.~Similarly for the other component,
\bea
p^y &=& \frac{a_0}{a}\bar p^y \nonumber\\
&=& \frac{a_0}{a} \left[ p^y(0) \pm \frac{g_D E^y_0}{k_P} 2\sin^2(\frac{k_P}{H_0} (\frac{a}{a_0} -1))\sin(k z) \right]\,.~~~~~
\eea
Therefore, the physical momentum is always redshifted with the expansion.

As an example in~\fref{plotARI} we show the time evolution after inflation of a couple of modes (the real and imaginary components) of  the vector potential and its time derivative.~We have used a quartic chaotic potential $V=\lambda \phi/4$, with $\lambda=10^{-14}$, and $\alpha m_P/f=7.5$.~The integration starts before inflation ends, but we only show the evolution from $\epsilon_H=1$ ($N_e=0$) onwards.~In the plot it is indicated the value of the comoving momentum in units of the $a_eH_e$, where $H_e$ is the value of the Hubble parameter at the end of inflation. 

%
\begin{figure}[tbh]
\begin{center}
\includegraphics[scale=0.29]{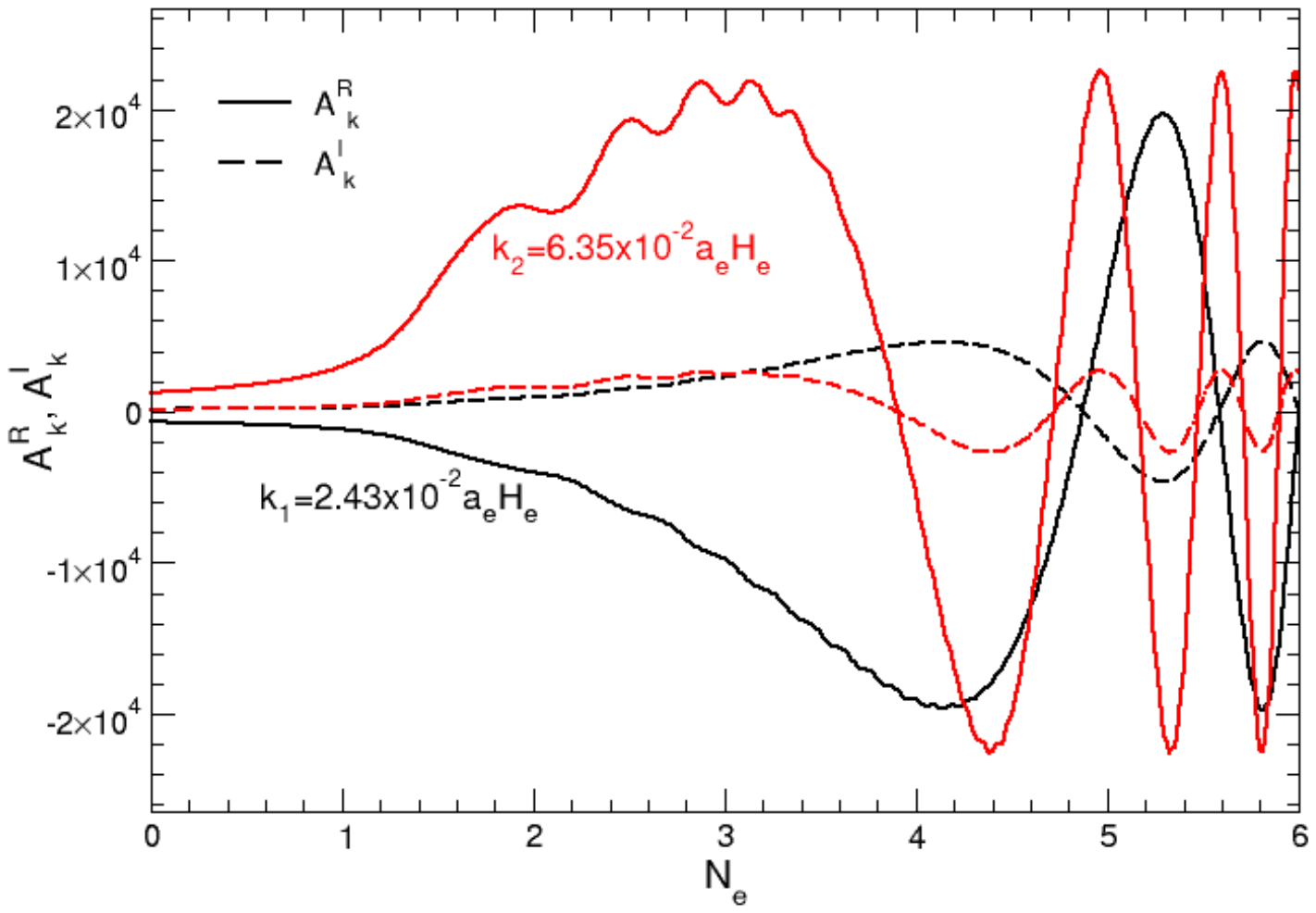}
\includegraphics[scale=0.29]{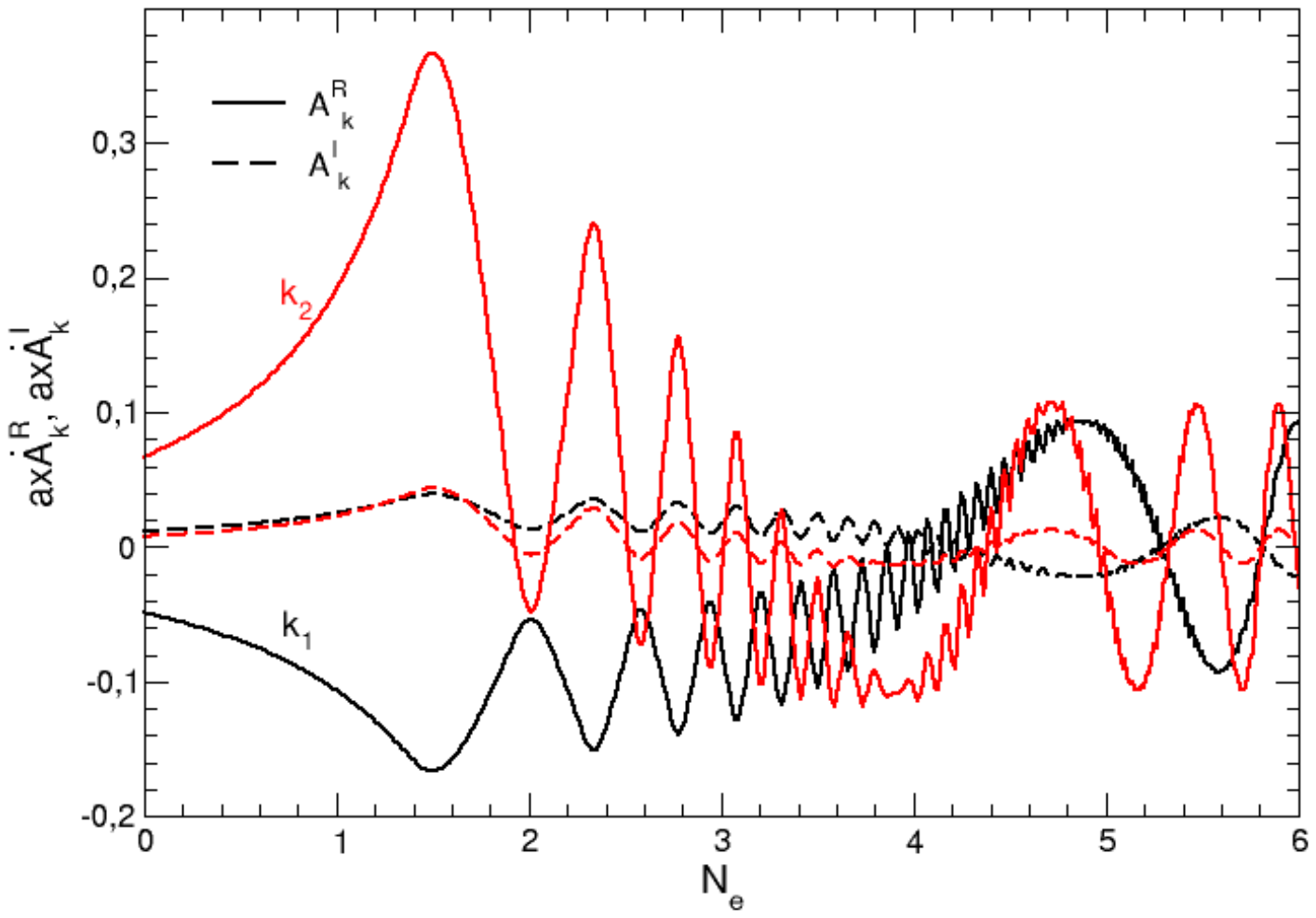}
\end{center}
\caption{Examples of the time evolution of the real $A^R_k$ and imaginary 
component $A^I_k$ of the vector potential (top panel) and its time
derivative (bottom panel), after inflation ends at $\epsilon_H=1$
($N_e=0$) onwards.~The values of the comoving momentum in units of
$a_eH_e$ are indicated in the plot, $H_e$ being the value of the Hubble
parameter at the end of inflation.~The values of the vector fields and
their derivate have been normalized by their initial values.
}
\label{fig:plotARI}
\end{figure}
%

\end{section}

\bibliographystyle{apsrev}
\bibliography{DarkSchwingerRefs}

\begin{thebibliography}{50}
\expandafter\ifx\csname natexlab\endcsname\relax\def\natexlab#1{#1}\fi
\expandafter\ifx\csname bibnamefont\endcsname\relax
  \def\bibnamefont#1{#1}\fi
\expandafter\ifx\csname bibfnamefont\endcsname\relax
  \def\bibfnamefont#1{#1}\fi
\expandafter\ifx\csname citenamefont\endcsname\relax
  \def\citenamefont#1{#1}\fi
\expandafter\ifx\csname url\endcsname\relax
  \def\url#1{\texttt{#1}}\fi
\expandafter\ifx\csname urlprefix\endcsname\relax\def\urlprefix{URL }\fi
\providecommand{\bibinfo}[2]{#2}
\providecommand{\eprint}[2][]{\url{#2}}

\bibitem[{\citenamefont{Bertone and Tait}(2018)}]{Bertone:2018xtm}
\bibinfo{author}{\bibfnamefont{G.}~\bibnamefont{Bertone}} \bibnamefont{and}
  \bibinfo{author}{\bibfnamefont{M.~P.} \bibnamefont{Tait},
  \bibfnamefont{Tim}}, \bibinfo{journal}{Nature}
  \textbf{\bibinfo{volume}{562}}, \bibinfo{pages}{51} (\bibinfo{year}{2018}),
  \eprint{1810.01668}.

\bibitem[{\citenamefont{Graham et~al.}(2016)\citenamefont{Graham, Mardon, and
  Rajendran}}]{Graham:2015rva}
\bibinfo{author}{\bibfnamefont{P.~W.} \bibnamefont{Graham}},
  \bibinfo{author}{\bibfnamefont{J.}~\bibnamefont{Mardon}}, \bibnamefont{and}
  \bibinfo{author}{\bibfnamefont{S.}~\bibnamefont{Rajendran}},
  \bibinfo{journal}{Phys. Rev.} \textbf{\bibinfo{volume}{D93}},
  \bibinfo{pages}{103520} (\bibinfo{year}{2016}), \eprint{1504.02102}.

\bibitem[{\citenamefont{Agrawal et~al.}(2020)\citenamefont{Agrawal, Kitajima,
  Reece, Sekiguchi, and Takahashi}}]{Agrawal:2018vin}
\bibinfo{author}{\bibfnamefont{P.}~\bibnamefont{Agrawal}},
  \bibinfo{author}{\bibfnamefont{N.}~\bibnamefont{Kitajima}},
  \bibinfo{author}{\bibfnamefont{M.}~\bibnamefont{Reece}},
  \bibinfo{author}{\bibfnamefont{T.}~\bibnamefont{Sekiguchi}},
  \bibnamefont{and}
  \bibinfo{author}{\bibfnamefont{F.}~\bibnamefont{Takahashi}},
  \bibinfo{journal}{Phys. Lett.} \textbf{\bibinfo{volume}{B801}},
  \bibinfo{pages}{135136} (\bibinfo{year}{2020}), \eprint{1810.07188}.

\bibitem[{\citenamefont{Dror et~al.}(2019)\citenamefont{Dror, Harigaya, and
  Narayan}}]{Dror:2018pdh}
\bibinfo{author}{\bibfnamefont{J.~A.} \bibnamefont{Dror}},
  \bibinfo{author}{\bibfnamefont{K.}~\bibnamefont{Harigaya}}, \bibnamefont{and}
  \bibinfo{author}{\bibfnamefont{V.}~\bibnamefont{Narayan}},
  \bibinfo{journal}{Phys. Rev.} \textbf{\bibinfo{volume}{D99}},
  \bibinfo{pages}{035036} (\bibinfo{year}{2019}), \eprint{1810.07195}.

\bibitem[{\citenamefont{Co et~al.}(2019)\citenamefont{Co, Pierce, Zhang, and
  Zhao}}]{Co:2018lka}
\bibinfo{author}{\bibfnamefont{R.~T.} \bibnamefont{Co}},
  \bibinfo{author}{\bibfnamefont{A.}~\bibnamefont{Pierce}},
  \bibinfo{author}{\bibfnamefont{Z.}~\bibnamefont{Zhang}}, \bibnamefont{and}
  \bibinfo{author}{\bibfnamefont{Y.}~\bibnamefont{Zhao}},
  \bibinfo{journal}{Phys. Rev.} \textbf{\bibinfo{volume}{D99}},
  \bibinfo{pages}{075002} (\bibinfo{year}{2019}), \eprint{1810.07196}.

\bibitem[{\citenamefont{Bastero-Gil et~al.}(2019)\citenamefont{Bastero-Gil,
  Santiago, Ubaldi, and Vega-Morales}}]{Bastero-Gil:2018uel}
\bibinfo{author}{\bibfnamefont{M.}~\bibnamefont{Bastero-Gil}},
  \bibinfo{author}{\bibfnamefont{J.}~\bibnamefont{Santiago}},
  \bibinfo{author}{\bibfnamefont{L.}~\bibnamefont{Ubaldi}}, \bibnamefont{and}
  \bibinfo{author}{\bibfnamefont{R.}~\bibnamefont{Vega-Morales}},
  \bibinfo{journal}{JCAP} \textbf{\bibinfo{volume}{1904}}, \bibinfo{pages}{015}
  (\bibinfo{year}{2019}), \eprint{1810.07208}.

\bibitem[{\citenamefont{Long and Wang}(2019)}]{Long:2019lwl}
\bibinfo{author}{\bibfnamefont{A.~J.} \bibnamefont{Long}} \bibnamefont{and}
  \bibinfo{author}{\bibfnamefont{L.-T.} \bibnamefont{Wang}},
  \bibinfo{journal}{Phys. Rev. D} \textbf{\bibinfo{volume}{99}},
  \bibinfo{pages}{063529} (\bibinfo{year}{2019}), \eprint{1901.03312}.

\bibitem[{\citenamefont{McDermott and Witte}(2020)}]{McDermott:2019lch}
\bibinfo{author}{\bibfnamefont{S.~D.} \bibnamefont{McDermott}}
  \bibnamefont{and} \bibinfo{author}{\bibfnamefont{S.~J.} \bibnamefont{Witte}},
  \bibinfo{journal}{Phys. Rev. D} \textbf{\bibinfo{volume}{101}},
  \bibinfo{pages}{063030} (\bibinfo{year}{2020}), \eprint{1911.05086}.

\bibitem[{\citenamefont{Nakai et~al.}(2020)\citenamefont{Nakai, Namba, and
  Wang}}]{Nakai:2020cfw}
\bibinfo{author}{\bibfnamefont{Y.}~\bibnamefont{Nakai}},
  \bibinfo{author}{\bibfnamefont{R.}~\bibnamefont{Namba}}, \bibnamefont{and}
  \bibinfo{author}{\bibfnamefont{Z.}~\bibnamefont{Wang}}
  (\bibinfo{year}{2020}), \eprint{2004.10743}.

\bibitem[{\citenamefont{Ahmed et~al.}(2020)\citenamefont{Ahmed, Grzadkowski,
  and Socha}}]{Ahmed:2020fhc}
\bibinfo{author}{\bibfnamefont{A.}~\bibnamefont{Ahmed}},
  \bibinfo{author}{\bibfnamefont{B.}~\bibnamefont{Grzadkowski}},
  \bibnamefont{and} \bibinfo{author}{\bibfnamefont{A.}~\bibnamefont{Socha}},
  \bibinfo{journal}{JHEP} \textbf{\bibinfo{volume}{08}}, \bibinfo{pages}{059}
  (\bibinfo{year}{2020}), \eprint{2005.01766}.

\bibitem[{\citenamefont{Kolb and Long}(2021)}]{Kolb:2020fwh}
\bibinfo{author}{\bibfnamefont{E.~W.} \bibnamefont{Kolb}} \bibnamefont{and}
  \bibinfo{author}{\bibfnamefont{A.~J.} \bibnamefont{Long}},
  \bibinfo{journal}{JHEP} \textbf{\bibinfo{volume}{03}}, \bibinfo{pages}{283}
  (\bibinfo{year}{2021}), \eprint{2009.03828}.

\bibitem[{\citenamefont{Salehian et~al.}(2021)\citenamefont{Salehian, Gorji,
  Firouzjahi, and Mukohyama}}]{Salehian:2020asa}
\bibinfo{author}{\bibfnamefont{B.}~\bibnamefont{Salehian}},
  \bibinfo{author}{\bibfnamefont{M.~A.} \bibnamefont{Gorji}},
  \bibinfo{author}{\bibfnamefont{H.}~\bibnamefont{Firouzjahi}},
  \bibnamefont{and}
  \bibinfo{author}{\bibfnamefont{S.}~\bibnamefont{Mukohyama}},
  \bibinfo{journal}{Phys. Rev. D} \textbf{\bibinfo{volume}{103}},
  \bibinfo{pages}{063526} (\bibinfo{year}{2021}), \eprint{2010.04491}.

\bibitem[{\citenamefont{Bastero-Gil et~al.}(2022)\citenamefont{Bastero-Gil,
  Santiago, Vega-Morales, and Ubaldi}}]{Bastero-Gil:2021wsf}
\bibinfo{author}{\bibfnamefont{M.}~\bibnamefont{Bastero-Gil}},
  \bibinfo{author}{\bibfnamefont{J.}~\bibnamefont{Santiago}},
  \bibinfo{author}{\bibfnamefont{R.}~\bibnamefont{Vega-Morales}},
  \bibnamefont{and} \bibinfo{author}{\bibfnamefont{L.}~\bibnamefont{Ubaldi}},
  \bibinfo{journal}{JCAP} \textbf{\bibinfo{volume}{02}}, \bibinfo{pages}{015}
  (\bibinfo{year}{2022}), \eprint{2103.12145}.

\bibitem[{\citenamefont{Gorghetto et~al.}(2022)\citenamefont{Gorghetto, Hardy,
  March-Russell, Song, and West}}]{Gorghetto:2022sue}
\bibinfo{author}{\bibfnamefont{M.}~\bibnamefont{Gorghetto}},
  \bibinfo{author}{\bibfnamefont{E.}~\bibnamefont{Hardy}},
  \bibinfo{author}{\bibfnamefont{J.}~\bibnamefont{March-Russell}},
  \bibinfo{author}{\bibfnamefont{N.}~\bibnamefont{Song}}, \bibnamefont{and}
  \bibinfo{author}{\bibfnamefont{S.~M.} \bibnamefont{West}},
  \bibinfo{journal}{JCAP} \textbf{\bibinfo{volume}{08}}, \bibinfo{pages}{018}
  (\bibinfo{year}{2022}), \eprint{2203.10100}.

\bibitem[{\citenamefont{Sato et~al.}(2022)\citenamefont{Sato, Takahashi, and
  Yamada}}]{Sato:2022jya}
\bibinfo{author}{\bibfnamefont{T.}~\bibnamefont{Sato}},
  \bibinfo{author}{\bibfnamefont{F.}~\bibnamefont{Takahashi}},
  \bibnamefont{and} \bibinfo{author}{\bibfnamefont{M.}~\bibnamefont{Yamada}},
  \bibinfo{journal}{JCAP} \textbf{\bibinfo{volume}{08}}, \bibinfo{pages}{022}
  (\bibinfo{year}{2022}), \eprint{2204.11896}.

\bibitem[{\citenamefont{Redi and Tesi}(2022)}]{Redi:2022zkt}
\bibinfo{author}{\bibfnamefont{M.}~\bibnamefont{Redi}} \bibnamefont{and}
  \bibinfo{author}{\bibfnamefont{A.}~\bibnamefont{Tesi}},
  \bibinfo{journal}{JHEP} \textbf{\bibinfo{volume}{10}}, \bibinfo{pages}{167}
  (\bibinfo{year}{2022}), \eprint{2204.14274}.

\bibitem[{\citenamefont{Barrie}(2022)}]{Barrie:2022mma}
\bibinfo{author}{\bibfnamefont{N.~D.} \bibnamefont{Barrie}}
  (\bibinfo{year}{2022}), \eprint{2211.03902}.

\bibitem[{\citenamefont{East and Huang}(2022)}]{East:2022rsi}
\bibinfo{author}{\bibfnamefont{W.~E.} \bibnamefont{East}} \bibnamefont{and}
  \bibinfo{author}{\bibfnamefont{J.}~\bibnamefont{Huang}},
  \bibinfo{journal}{JHEP} \textbf{\bibinfo{volume}{12}}, \bibinfo{pages}{089}
  (\bibinfo{year}{2022}), \eprint{2206.12432}.

\bibitem[{\citenamefont{Cyncynates and Weiner}(2023)}]{Cyncynates:2023zwj}
\bibinfo{author}{\bibfnamefont{D.}~\bibnamefont{Cyncynates}} \bibnamefont{and}
  \bibinfo{author}{\bibfnamefont{Z.~J.} \bibnamefont{Weiner}}
  (\bibinfo{year}{2023}), \eprint{2310.18397}.

\bibitem[{\citenamefont{Arvanitaki et~al.}(2021)\citenamefont{Arvanitaki,
  Dimopoulos, Galanis, Racco, Simon, and Thompson}}]{Arvanitaki:2021qlj}
\bibinfo{author}{\bibfnamefont{A.}~\bibnamefont{Arvanitaki}},
  \bibinfo{author}{\bibfnamefont{S.}~\bibnamefont{Dimopoulos}},
  \bibinfo{author}{\bibfnamefont{M.}~\bibnamefont{Galanis}},
  \bibinfo{author}{\bibfnamefont{D.}~\bibnamefont{Racco}},
  \bibinfo{author}{\bibfnamefont{O.}~\bibnamefont{Simon}}, \bibnamefont{and}
  \bibinfo{author}{\bibfnamefont{J.~O.} \bibnamefont{Thompson}},
  \bibinfo{journal}{JHEP} \textbf{\bibinfo{volume}{11}}, \bibinfo{pages}{106}
  (\bibinfo{year}{2021}), \eprint{2108.04823}.

\bibitem[{\citenamefont{Fröb et~al.}(2014)\citenamefont{Fröb, Garriga, Kanno,
  Sasaki, Soda, Tanaka, and Vilenkin}}]{Frob:2014zka}
\bibinfo{author}{\bibfnamefont{M.~B.} \bibnamefont{Fröb}},
  \bibinfo{author}{\bibfnamefont{J.}~\bibnamefont{Garriga}},
  \bibinfo{author}{\bibfnamefont{S.}~\bibnamefont{Kanno}},
  \bibinfo{author}{\bibfnamefont{M.}~\bibnamefont{Sasaki}},
  \bibinfo{author}{\bibfnamefont{J.}~\bibnamefont{Soda}},
  \bibinfo{author}{\bibfnamefont{T.}~\bibnamefont{Tanaka}}, \bibnamefont{and}
  \bibinfo{author}{\bibfnamefont{A.}~\bibnamefont{Vilenkin}},
  \bibinfo{journal}{JCAP} \textbf{\bibinfo{volume}{1404}}, \bibinfo{pages}{009}
  (\bibinfo{year}{2014}), \eprint{1401.4137}.

\bibitem[{\citenamefont{Kobayashi and Afshordi}(2014)}]{Kobayashi:2014zza}
\bibinfo{author}{\bibfnamefont{T.}~\bibnamefont{Kobayashi}} \bibnamefont{and}
  \bibinfo{author}{\bibfnamefont{N.}~\bibnamefont{Afshordi}},
  \bibinfo{journal}{JHEP} \textbf{\bibinfo{volume}{10}}, \bibinfo{pages}{166}
  (\bibinfo{year}{2014}), \eprint{1408.4141}.

\bibitem[{\citenamefont{Hayashinaka et~al.}(2016)\citenamefont{Hayashinaka,
  Fujita, and Yokoyama}}]{Hayashinaka:2016qqn}
\bibinfo{author}{\bibfnamefont{T.}~\bibnamefont{Hayashinaka}},
  \bibinfo{author}{\bibfnamefont{T.}~\bibnamefont{Fujita}}, \bibnamefont{and}
  \bibinfo{author}{\bibfnamefont{J.}~\bibnamefont{Yokoyama}},
  \bibinfo{journal}{JCAP} \textbf{\bibinfo{volume}{1607}}, \bibinfo{pages}{010}
  (\bibinfo{year}{2016}), \eprint{1603.04165}.

\bibitem[{\citenamefont{Banyeres et~al.}(2018)\citenamefont{Banyeres,
  Domènech, and Garriga}}]{Banyeres:2018aax}
\bibinfo{author}{\bibfnamefont{M.}~\bibnamefont{Banyeres}},
  \bibinfo{author}{\bibfnamefont{G.}~\bibnamefont{Domènech}},
  \bibnamefont{and} \bibinfo{author}{\bibfnamefont{J.}~\bibnamefont{Garriga}},
  \bibinfo{journal}{JCAP} \textbf{\bibinfo{volume}{1810}}, \bibinfo{pages}{023}
  (\bibinfo{year}{2018}), \eprint{1809.08977}.

\bibitem[{\citenamefont{Domcke and Mukaida}(2018)}]{Domcke:2018eki}
\bibinfo{author}{\bibfnamefont{V.}~\bibnamefont{Domcke}} \bibnamefont{and}
  \bibinfo{author}{\bibfnamefont{K.}~\bibnamefont{Mukaida}},
  \bibinfo{journal}{JCAP} \textbf{\bibinfo{volume}{11}}, \bibinfo{pages}{020}
  (\bibinfo{year}{2018}), \eprint{1806.08769}.

\bibitem[{\citenamefont{Stahl}(2019)}]{Stahl:2018idd}
\bibinfo{author}{\bibfnamefont{C.}~\bibnamefont{Stahl}},
  \bibinfo{journal}{Nucl. Phys.} \textbf{\bibinfo{volume}{B939}},
  \bibinfo{pages}{95} (\bibinfo{year}{2019}), \eprint{1806.06692}.

\bibitem[{\citenamefont{Sobol et~al.}(2019)\citenamefont{Sobol, Gorbar, and
  Vilchinskii}}]{Sobol:2019xls}
\bibinfo{author}{\bibfnamefont{O.~O.} \bibnamefont{Sobol}},
  \bibinfo{author}{\bibfnamefont{E.~V.} \bibnamefont{Gorbar}},
  \bibnamefont{and} \bibinfo{author}{\bibfnamefont{S.~I.}
  \bibnamefont{Vilchinskii}}, \bibinfo{journal}{Phys. Rev.}
  \textbf{\bibinfo{volume}{D100}}, \bibinfo{pages}{063523}
  (\bibinfo{year}{2019}), \eprint{1907.10443}.

\bibitem[{\citenamefont{Domcke et~al.}(2020)\citenamefont{Domcke, Ema, and
  Mukaida}}]{Domcke:2019qmm}
\bibinfo{author}{\bibfnamefont{V.}~\bibnamefont{Domcke}},
  \bibinfo{author}{\bibfnamefont{Y.}~\bibnamefont{Ema}}, \bibnamefont{and}
  \bibinfo{author}{\bibfnamefont{K.}~\bibnamefont{Mukaida}},
  \bibinfo{journal}{JHEP} \textbf{\bibinfo{volume}{02}}, \bibinfo{pages}{055}
  (\bibinfo{year}{2020}), \eprint{1910.01205}.

\bibitem[{\citenamefont{Sobol et~al.}(2020)\citenamefont{Sobol, Gorbar, Momot,
  and Vilchinskii}}]{Sobol:2020frh}
\bibinfo{author}{\bibfnamefont{O.~O.} \bibnamefont{Sobol}},
  \bibinfo{author}{\bibfnamefont{E.~V.} \bibnamefont{Gorbar}},
  \bibinfo{author}{\bibfnamefont{A.~I.} \bibnamefont{Momot}}, \bibnamefont{and}
  \bibinfo{author}{\bibfnamefont{S.~I.} \bibnamefont{Vilchinskii}}
  (\bibinfo{year}{2020}), \eprint{2004.12664}.

\bibitem[{\citenamefont{Ford}(1987)}]{Ford:1986sy}
\bibinfo{author}{\bibfnamefont{L.~H.} \bibnamefont{Ford}},
  \bibinfo{journal}{Phys. Rev. D} \textbf{\bibinfo{volume}{35}},
  \bibinfo{pages}{2955} (\bibinfo{year}{1987}).

\bibitem[{\citenamefont{Kuzmin and Tkachev}(1999)}]{Kuzmin:1998kk}
\bibinfo{author}{\bibfnamefont{V.}~\bibnamefont{Kuzmin}} \bibnamefont{and}
  \bibinfo{author}{\bibfnamefont{I.}~\bibnamefont{Tkachev}},
  \bibinfo{journal}{Phys. Rev. D} \textbf{\bibinfo{volume}{59}},
  \bibinfo{pages}{123006} (\bibinfo{year}{1999}), \eprint{hep-ph/9809547}.

\bibitem[{\citenamefont{Bastero-Gil et~al.}(2023)\citenamefont{Bastero-Gil,
  Ferraz, Ubaldi, and Vega-Morales}}]{Bastero-Gil:2023htv}
\bibinfo{author}{\bibfnamefont{M.}~\bibnamefont{Bastero-Gil}},
  \bibinfo{author}{\bibfnamefont{P.~B.} \bibnamefont{Ferraz}},
  \bibinfo{author}{\bibfnamefont{L.}~\bibnamefont{Ubaldi}}, \bibnamefont{and}
  \bibinfo{author}{\bibfnamefont{R.}~\bibnamefont{Vega-Morales}}
  (\bibinfo{year}{2023}), \eprint{2311.09475}.

\bibitem[{\citenamefont{Shakeri et~al.}(2019)\citenamefont{Shakeri, Gorji, and
  Firouzjahi}}]{Shakeri:2019mnt}
\bibinfo{author}{\bibfnamefont{S.}~\bibnamefont{Shakeri}},
  \bibinfo{author}{\bibfnamefont{M.~A.} \bibnamefont{Gorji}}, \bibnamefont{and}
  \bibinfo{author}{\bibfnamefont{H.}~\bibnamefont{Firouzjahi}},
  \bibinfo{journal}{Phys. Rev. D} \textbf{\bibinfo{volume}{99}},
  \bibinfo{pages}{103525} (\bibinfo{year}{2019}), \eprint{1903.05310}.

\bibitem[{\citenamefont{Nakayama}(2020)}]{Nakayama:2020rka}
\bibinfo{author}{\bibfnamefont{K.}~\bibnamefont{Nakayama}}
  (\bibinfo{year}{2020}), \eprint{2004.10036}.

\bibitem[{\citenamefont{Bastero-Gil and Manso}(2023)}]{Bastero-Gil:2022fme}
\bibinfo{author}{\bibfnamefont{M.}~\bibnamefont{Bastero-Gil}} \bibnamefont{and}
  \bibinfo{author}{\bibfnamefont{A.~T.} \bibnamefont{Manso}},
  \bibinfo{journal}{JCAP} \textbf{\bibinfo{volume}{08}}, \bibinfo{pages}{001}
  (\bibinfo{year}{2023}), \eprint{2209.15572}.

\bibitem[{\citenamefont{Meerburg and Pajer}(2013)}]{Meer}
\bibinfo{author}{\bibfnamefont{P.~D.} \bibnamefont{Meerburg}} \bibnamefont{and}
  \bibinfo{author}{\bibfnamefont{E.}~\bibnamefont{Pajer}},
  \bibinfo{journal}{JCAP} \textbf{\bibinfo{volume}{1302}}, \bibinfo{pages}{017}
  (\bibinfo{year}{2013}), \eprint{1203.6076}.

\bibitem[{\citenamefont{Anber and Sorbo}(2010)}]{Anber:2009ua}
\bibinfo{author}{\bibfnamefont{M.~M.} \bibnamefont{Anber}} \bibnamefont{and}
  \bibinfo{author}{\bibfnamefont{L.}~\bibnamefont{Sorbo}},
  \bibinfo{journal}{Phys. Rev. D} \textbf{\bibinfo{volume}{81}},
  \bibinfo{pages}{043534} (\bibinfo{year}{2010}), \eprint{0908.4089}.

\bibitem[{\citenamefont{Barnaby et~al.}(2012)\citenamefont{Barnaby, Pajer, and
  Peloso}}]{Barnaby:2011qe}
\bibinfo{author}{\bibfnamefont{N.}~\bibnamefont{Barnaby}},
  \bibinfo{author}{\bibfnamefont{E.}~\bibnamefont{Pajer}}, \bibnamefont{and}
  \bibinfo{author}{\bibfnamefont{M.}~\bibnamefont{Peloso}},
  \bibinfo{journal}{Phys. Rev. D} \textbf{\bibinfo{volume}{85}},
  \bibinfo{pages}{023525} (\bibinfo{year}{2012}), \eprint{1110.3327}.

\bibitem[{\citenamefont{Fujita et~al.}(2022)\citenamefont{Fujita, Kume,
  Mukaida, and Tada}}]{Fujita:2022fwc}
\bibinfo{author}{\bibfnamefont{T.}~\bibnamefont{Fujita}},
  \bibinfo{author}{\bibfnamefont{J.}~\bibnamefont{Kume}},
  \bibinfo{author}{\bibfnamefont{K.}~\bibnamefont{Mukaida}}, \bibnamefont{and}
  \bibinfo{author}{\bibfnamefont{Y.}~\bibnamefont{Tada}},
  \bibinfo{journal}{JCAP} \textbf{\bibinfo{volume}{09}}, \bibinfo{pages}{023}
  (\bibinfo{year}{2022}), \eprint{2204.01180}.

\bibitem[{\citenamefont{Lasenby}(2020)}]{Lasenby:2020rlf}
\bibinfo{author}{\bibfnamefont{R.}~\bibnamefont{Lasenby}},
  \bibinfo{journal}{JCAP} \textbf{\bibinfo{volume}{11}}, \bibinfo{pages}{034}
  (\bibinfo{year}{2020}), \eprint{2007.00667}.

\bibitem[{\citenamefont{Agrawal et~al.}(2017)\citenamefont{Agrawal, Cyr-Racine,
  Randall, and Scholtz}}]{Agrawal:2016quu}
\bibinfo{author}{\bibfnamefont{P.}~\bibnamefont{Agrawal}},
  \bibinfo{author}{\bibfnamefont{F.-Y.} \bibnamefont{Cyr-Racine}},
  \bibinfo{author}{\bibfnamefont{L.}~\bibnamefont{Randall}}, \bibnamefont{and}
  \bibinfo{author}{\bibfnamefont{J.}~\bibnamefont{Scholtz}},
  \bibinfo{journal}{JCAP} \textbf{\bibinfo{volume}{05}}, \bibinfo{pages}{022}
  (\bibinfo{year}{2017}), \eprint{1610.04611}.

\bibitem[{\citenamefont{Arkani-Hamed et~al.}(2007)\citenamefont{Arkani-Hamed,
  Motl, Nicolis, and Vafa}}]{Arkani-Hamed:2006emk}
\bibinfo{author}{\bibfnamefont{N.}~\bibnamefont{Arkani-Hamed}},
  \bibinfo{author}{\bibfnamefont{L.}~\bibnamefont{Motl}},
  \bibinfo{author}{\bibfnamefont{A.}~\bibnamefont{Nicolis}}, \bibnamefont{and}
  \bibinfo{author}{\bibfnamefont{C.}~\bibnamefont{Vafa}},
  \bibinfo{journal}{JHEP} \textbf{\bibinfo{volume}{06}}, \bibinfo{pages}{060}
  (\bibinfo{year}{2007}), \eprint{hep-th/0601001}.

\bibitem[{\citenamefont{Siemonsen et~al.}(2023)\citenamefont{Siemonsen,
  Mondino, Egana-Ugrinovic, Huang, Baryakhtar, and East}}]{Siemonsen:2022ivj}
\bibinfo{author}{\bibfnamefont{N.}~\bibnamefont{Siemonsen}},
  \bibinfo{author}{\bibfnamefont{C.}~\bibnamefont{Mondino}},
  \bibinfo{author}{\bibfnamefont{D.}~\bibnamefont{Egana-Ugrinovic}},
  \bibinfo{author}{\bibfnamefont{J.}~\bibnamefont{Huang}},
  \bibinfo{author}{\bibfnamefont{M.}~\bibnamefont{Baryakhtar}},
  \bibnamefont{and} \bibinfo{author}{\bibfnamefont{W.~E.} \bibnamefont{East}},
  \bibinfo{journal}{Phys. Rev. D} \textbf{\bibinfo{volume}{107}},
  \bibinfo{pages}{075025} (\bibinfo{year}{2023}), \eprint{2212.09772}.

\bibitem[{\citenamefont{Ade et~al.}(2016)}]{Ade:2015xua}
\bibinfo{author}{\bibfnamefont{P.~A.~R.} \bibnamefont{Ade}}
  \bibnamefont{et~al.} (\bibinfo{collaboration}{Planck}),
  \bibinfo{journal}{Astron. Astrophys.} \textbf{\bibinfo{volume}{594}},
  \bibinfo{pages}{A13} (\bibinfo{year}{2016}), \eprint{1502.01589}.

\bibitem[{\citenamefont{von Eckardstein et~al.}(2023)\citenamefont{von
  Eckardstein, Peloso, Schmitz, Sobol, and Sorbo}}]{vonEckardstein:2023gwk}
\bibinfo{author}{\bibfnamefont{R.}~\bibnamefont{von Eckardstein}},
  \bibinfo{author}{\bibfnamefont{M.}~\bibnamefont{Peloso}},
  \bibinfo{author}{\bibfnamefont{K.}~\bibnamefont{Schmitz}},
  \bibinfo{author}{\bibfnamefont{O.}~\bibnamefont{Sobol}}, \bibnamefont{and}
  \bibinfo{author}{\bibfnamefont{L.}~\bibnamefont{Sorbo}},
  \bibinfo{journal}{JHEP} \textbf{\bibinfo{volume}{11}}, \bibinfo{pages}{183}
  (\bibinfo{year}{2023}), \eprint{2309.04254}.

\bibitem[{\citenamefont{Di~Paolo et~al.}(2018)\citenamefont{Di~Paolo, Nesti,
  and Villante}}]{DiPaolo:2017geq}
\bibinfo{author}{\bibfnamefont{C.}~\bibnamefont{Di~Paolo}},
  \bibinfo{author}{\bibfnamefont{F.}~\bibnamefont{Nesti}}, \bibnamefont{and}
  \bibinfo{author}{\bibfnamefont{F.~L.} \bibnamefont{Villante}},
  \bibinfo{journal}{Mon. Not. Roy. Astron. Soc.}
  \textbf{\bibinfo{volume}{475}}, \bibinfo{pages}{5385} (\bibinfo{year}{2018}),
  \eprint{1704.06644}.

\bibitem[{\citenamefont{Alvey et~al.}(2021)\citenamefont{Alvey, Sabti, Tiki,
  Blas, Bondarenko, Boyarsky, Escudero, Fairbairn, Orkney, and
  Read}}]{Alvey:2020xsk}
\bibinfo{author}{\bibfnamefont{J.}~\bibnamefont{Alvey}},
  \bibinfo{author}{\bibfnamefont{N.}~\bibnamefont{Sabti}},
  \bibinfo{author}{\bibfnamefont{V.}~\bibnamefont{Tiki}},
  \bibinfo{author}{\bibfnamefont{D.}~\bibnamefont{Blas}},
  \bibinfo{author}{\bibfnamefont{K.}~\bibnamefont{Bondarenko}},
  \bibinfo{author}{\bibfnamefont{A.}~\bibnamefont{Boyarsky}},
  \bibinfo{author}{\bibfnamefont{M.}~\bibnamefont{Escudero}},
  \bibinfo{author}{\bibfnamefont{M.}~\bibnamefont{Fairbairn}},
  \bibinfo{author}{\bibfnamefont{M.}~\bibnamefont{Orkney}}, \bibnamefont{and}
  \bibinfo{author}{\bibfnamefont{J.~I.} \bibnamefont{Read}},
  \bibinfo{journal}{Mon. Not. Roy. Astron. Soc.}
  \textbf{\bibinfo{volume}{501}}, \bibinfo{pages}{1188} (\bibinfo{year}{2021}),
  \eprint{2010.03572}.

\bibitem[{\citenamefont{Archidiacono et~al.}(2022)\citenamefont{Archidiacono,
  Castorina, Redigolo, and Salvioni}}]{Archidiacono:2022iuu}
\bibinfo{author}{\bibfnamefont{M.}~\bibnamefont{Archidiacono}},
  \bibinfo{author}{\bibfnamefont{E.}~\bibnamefont{Castorina}},
  \bibinfo{author}{\bibfnamefont{D.}~\bibnamefont{Redigolo}}, \bibnamefont{and}
  \bibinfo{author}{\bibfnamefont{E.}~\bibnamefont{Salvioni}},
  \bibinfo{journal}{JCAP} \textbf{\bibinfo{volume}{10}}, \bibinfo{pages}{074}
  (\bibinfo{year}{2022}), \eprint{2204.08484}.

\bibitem[{\citenamefont{Bottaro
  et~al.}(2024{\natexlab{a}})\citenamefont{Bottaro, Castorina, Costa, Redigolo,
  and Salvioni}}]{Bottaro:2023wkd}
\bibinfo{author}{\bibfnamefont{S.}~\bibnamefont{Bottaro}},
  \bibinfo{author}{\bibfnamefont{E.}~\bibnamefont{Castorina}},
  \bibinfo{author}{\bibfnamefont{M.}~\bibnamefont{Costa}},
  \bibinfo{author}{\bibfnamefont{D.}~\bibnamefont{Redigolo}}, \bibnamefont{and}
  \bibinfo{author}{\bibfnamefont{E.}~\bibnamefont{Salvioni}},
  \bibinfo{journal}{Phys. Rev. Lett.} \textbf{\bibinfo{volume}{132}},
  \bibinfo{pages}{201002} (\bibinfo{year}{2024}{\natexlab{a}}),
  \eprint{2309.11496}.

\bibitem[{\citenamefont{Bottaro
  et~al.}(2024{\natexlab{b}})\citenamefont{Bottaro, Castorina, Costa, Redigolo,
  and Salvioni}}]{Bottaro:2024pcb}
\bibinfo{author}{\bibfnamefont{S.}~\bibnamefont{Bottaro}},
  \bibinfo{author}{\bibfnamefont{E.}~\bibnamefont{Castorina}},
  \bibinfo{author}{\bibfnamefont{M.}~\bibnamefont{Costa}},
  \bibinfo{author}{\bibfnamefont{D.}~\bibnamefont{Redigolo}}, \bibnamefont{and}
  \bibinfo{author}{\bibfnamefont{E.}~\bibnamefont{Salvioni}}
  (\bibinfo{year}{2024}{\natexlab{b}}), \eprint{2407.18252}.

\end{thebibliography}

\end{document}